\documentclass[prd,preprint,tightenlines,floatfix,
showpacs,preprintnumbers,nofootinbib,eqsecnum,superscriptaddress]{revtex4}

\usepackage[T1]{fontenc}		

\usepackage{amsmath,amsfonts,amssymb,amstext,mathrsfs}
\usepackage{mathpazo}

\usepackage[dvips]{graphicx}
\usepackage{epsf,float}
\usepackage{revsymb}

\usepackage{dcolumn}
\usepackage{braket}
\usepackage{color,xcolor}
\usepackage{graphicx}
\usepackage{subfigure}
\usepackage{multirow}
\usepackage{tabularx}
\usepackage{pstricks}
\usepackage[section]{placeins}
\usepackage{booktabs}
\usepackage{array}

\usepackage{hyperref}

\begin{document}

\begin{center}

{\bf\Large
%
%
%
%
%
%
%
%
%
%
Rapidity distributions of pions in p+p and Pb+Pb collisions at CERN SPS energies
}

\vspace{0.5cm}

Andrzej Rybicki$^1$,
Antoni Szczurek$^{1,2}$, 
Miros\l{}aw Kie\l{}bowicz$^1$,
Antoni Marcinek$^1$,
Vitalii Ozvenchuk$^1$,
{\L{}ukasz Rozp\l{}ochowski$^3$}

\vspace{0.6cm}

{\small $^1$~H.~Niewodnicza\'{n}ski Institute of Nuclear Physics, Polish 
 Academy of Sciences, Radzikowskiego 152, 31-342~Krak\'ow, Poland\\
 $^2$~University of Rzesz\'ow, Rejtana 16, 35-959 Rzesz\'ow, Poland\\
{$^3$~Jagiellonian University, Krak\'ow, Poland}}
\end {center}

\vspace*{0.0cm}

{\small\bf Abstract}\\

{\small
The centrality dependence of rapidity distributions of pions in Pb+Pb reactions can be understood by imposing local energy-momentum conservation
in the longitudinal ``fire-streaks'' of excited matter.
With no tuning nor adjustment to the 
experimental data, 
the rapidity distribution of pions produced by 
the fire-streak
which we
obtained from Pb+Pb collisions reproduces the shape of the experimental pion rapidity distribution in p+p interactions,
measured by the NA49 Collaboration
 at the same energy. 
The observed difference in the absolute normalization of this distribution can be explained by 
the difference in the 
overall
energy balance, induced by
baryon stopping and strangeness enhancement phenomena occurring in heavy ion collisions.
We estimate the latter effects using a collection of SPS 
 experimental data on $\pi^\pm$, $K^\pm$, net $p$, and $n$ production in p+p and Pb+Pb reactions.
Implications of the above findings are discussed.}

\section{Introduction}
\label{intro}

\begin{figure}[p]
\begin{center}
\hspace*{-0.1cm}\includegraphics[width=10cm]{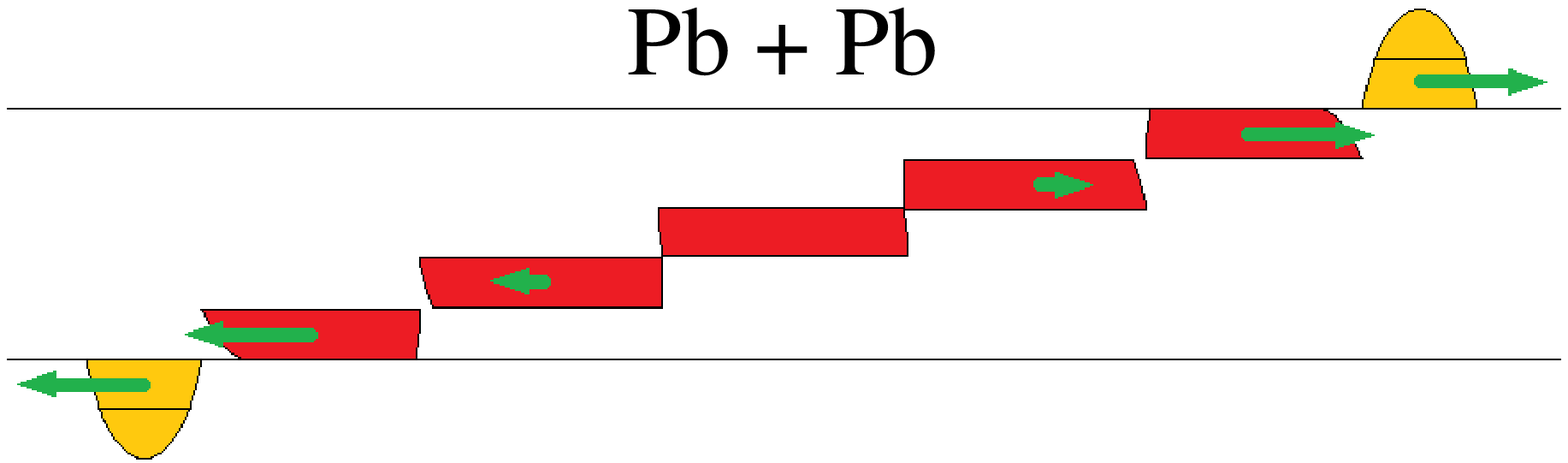}
{\caption\small 
A schematic picture of our model of Pb+Pb collisions~\cite{1}.
\label{fig1}}
\end{center}
\end{figure}

\begin{figure}[h]
\begin{center}
\vspace*{-0.1cm}
\vspace*{-0.5cm}
\hspace*{-4.7cm}
\includegraphics[width=15cm]{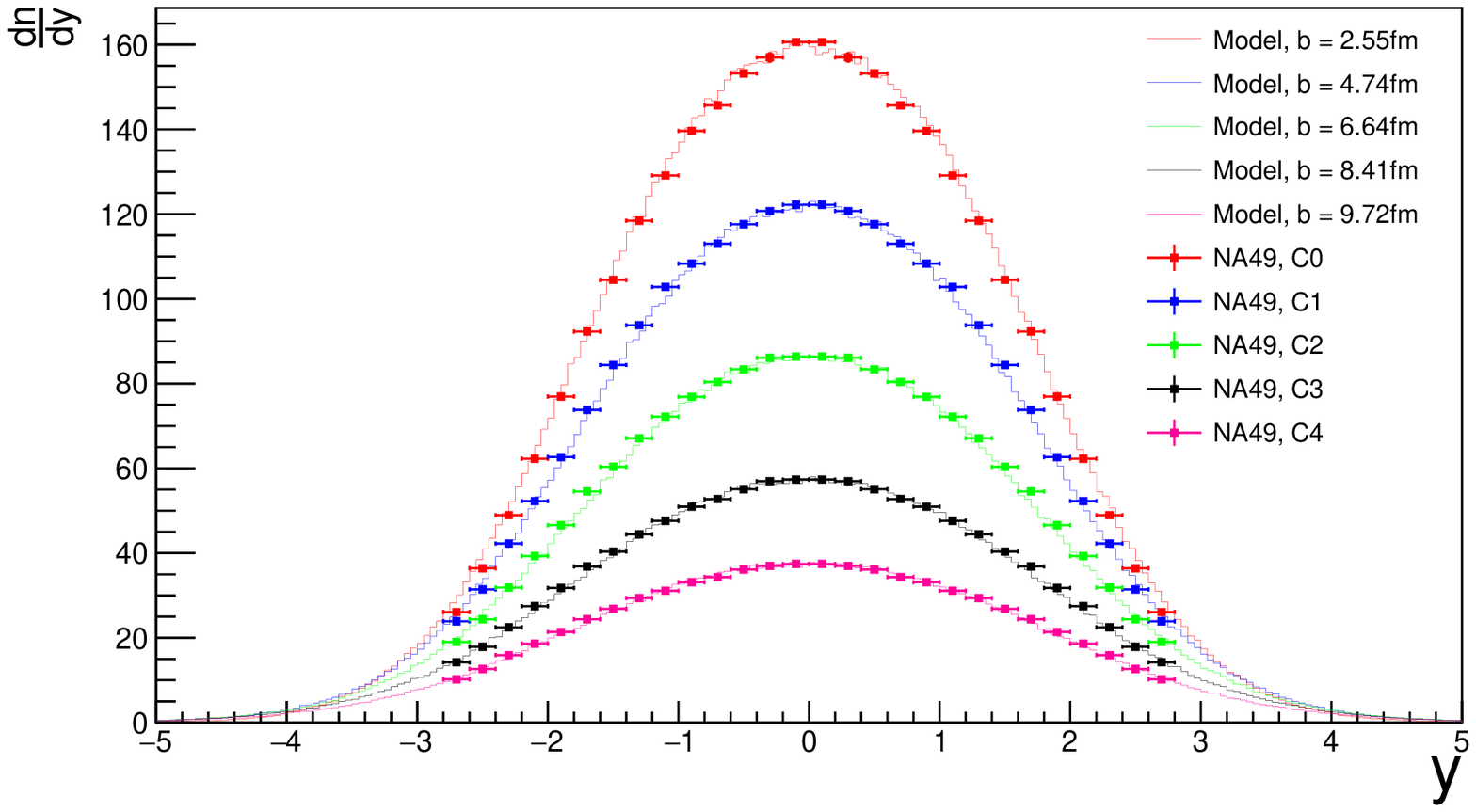}\vspace*{-5.5cm}
\hspace*{-0.7cm}
\hspace*{-0.5cm}
 \begin{picture}(10,10)
   \end{picture}
\includegraphics[width=6.5cm]{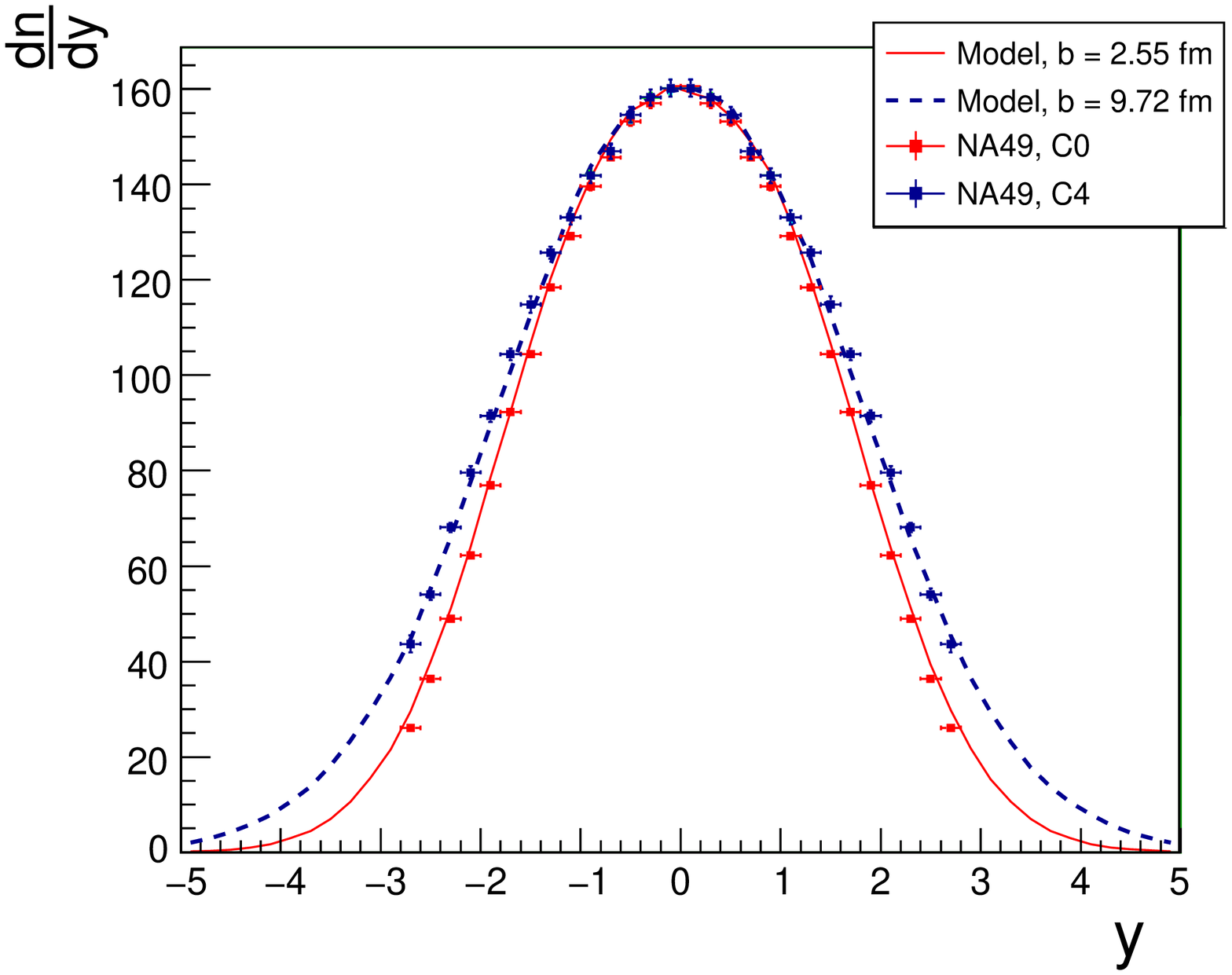}
\vspace*{.3cm}
 \begin{picture}(10,10)
 \put(-188,72){{\bf\Large{(a)}}}
 \put(187,72){{\bf\Large{(b)}}}
  \end{picture}
{\caption\small 
{(a) Rapidity distributions of $\pi^-$ mesons  in centrality selected Pb+Pb collisions at top SPS energy, $\sqrt{s_{NN}}=17.3$~GeV, together with our model calculations~\cite{1}, (b) change of width of the $\pi^-$ distribution from peripheral to central Pb+Pb collisions at $\sqrt{s_{NN}}=17.3$~GeV, and its description by our model~\cite{1}. In panel (b), for peripheral collisions, 
the
experimental data and model curves have been scaled up to fit the same maximum as for central collisions.}
\label{fig2}}
\vspace*{0.6cm}
\end{center}
\end{figure}

\begin{figure}[h]
\begin{center}
\vspace*{-0.8cm}
\hspace*{-0.7cm}
\vspace*{-0.5cm}
\includegraphics[width=6.4cm,height=5.3cm]{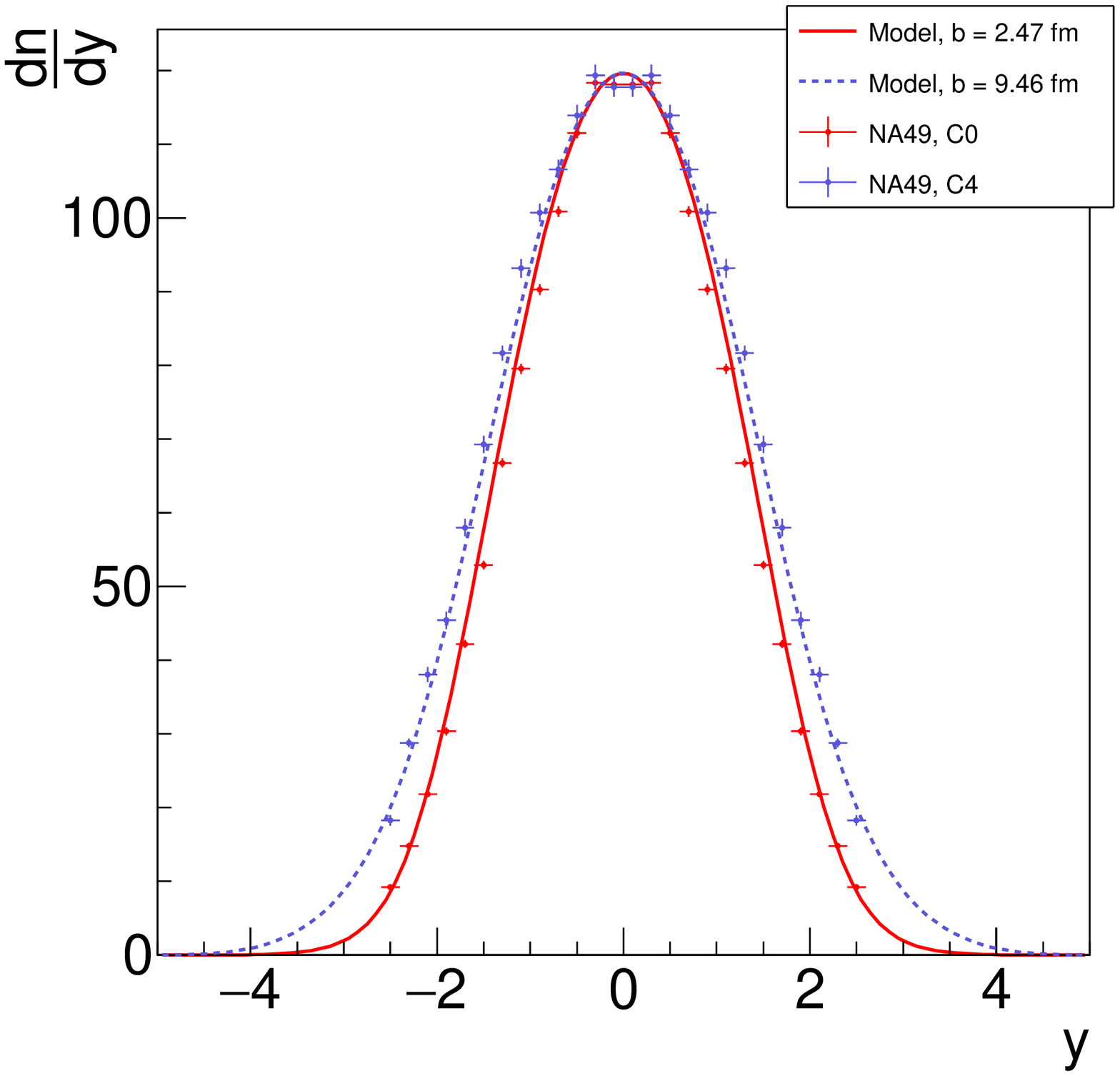}
{\caption\small 
{Change of width of the $\pi^-$ rapidity distribution from peripheral to central Pb+Pb collisions at 
the energy
$\sqrt{s_{NN}}=8.8$~GeV, and its description by our model~\cite{1}. For peripheral collisions, the
experimental data and model curves have been scaled up to fit the same maximum as for central collisions. The experimental data points come from the NA49 experiment~\cite{2.5na49}.}
\label{fig2-40}}
\end{center}
\end{figure}

In our recent paper on the implications of energy and momentum ($E$$-$$\vec{p}$)
conservation for heavy ion collisions at CERN SPS energies~\cite{1} we formulated a simple model for the longitudinal evolution of the participant system. This model, 
with some degree of similarity to the fire-streak approach of Refs~\cite{2}, assumed local 
$E$$-$$\vec{p}$
conservation 
in the plane perpendicular to the collision axis
and consequently, formation and independent fragmentation of finite volumes of 
excited 
primordial
matter (``fire-streaks'') into finite state particles. 
The kinematical characteristics (rapidity, invariant mass) of the fire-streaks were directly given by 
the
$E$$-$$\vec{p}$
conservation.
We did not address the exact physical nature of the fire-streaks although to think about color string conglomerates or initial volume elements of 
quark-gluon plasma
would not seem unnatural.
With a simple, 
three-parameter
fire-streak
fragmentation function ensuring energy conservation, our model provided a surprisingly good description of the whole centrality dependence of negative pion $dn/dy$ 
distributions
 in Pb+Pb reactions at $\sqrt{s_{NN}}=17.3$~GeV, measured by the NA49 experiment~\cite{2.5na49}. A reminder of the model is presented in Fig.~\ref{fig1}, while a compilation of results is shown in Fig.~\ref{fig2}.
It is noticeable that the model explains {\em both} the 
evolution of 
absolutely normalized
$\pi^-$ yields and of the distribution's shape as a function of centrality.
%
%
{In Fig.~\ref{fig2-40} we present the result of a first test of our model 
for Pb+Pb collisions at a lower SPS energy, $\sqrt{s_{NN}}=8.8$~GeV. The overall similarity of this result to that shown in Fig.~\ref{fig2}(b) suggests the applicability of our 
model to pion production in some extended range of collision energy, 8.8-17.3 GeV at the least.}

We interpreted the success of our simple model as a hint that
energy-momentum conservation indeed 
plays  a dominant
role in the longitudinal evolution of 
the system
created 
in A+A collisions,
at SPS energy.
Now we
wish to compare the results of our work on Pb+Pb collisions to more elementary p+p reactions.
The question whether the non-perturbative dynamical mechanisms governing the latter are qualitatively similar or different from
those in heavy ion reactions is a long-standing one. Evident differences on the quantitative level, including in particular the enhancement of strangeness production  and its energy dependence~\cite{3-40gev}, were interpreted as onset of deconfinement and transition to quark-gluon plasma~\cite{rafelski,smes}. On the other hand, qualitative similarities between p+p and Pb+Pb reactions at SPS~\cite{aduszkiewicz} and LHC energies~\cite{6-nature2017} still constitute a challenge for phenomenological models (see, e.g.,~\cite{7-ozvenchuk}). 
We find it therefore a key question to verify how our simple energy-momentum conservation picture in A+A reactions compares to proton-proton collisions.
%

This paper is organized as follows. In section~\ref{II} we remind the 
basic
formulae defining our fire-streak 
fragmentation function.
%
A comparison between 
the latter and p+p data from the NA49 Collaboration
is made in section~\ref{III}. 
The problem of isospin differences between p+p and Pb+Pb collisions is addressed in section~\ref{III.5}. 
Section~\ref{IV} includes the analysis of normalization.
The implications of our study are discussed in section~\ref{V} and the summary is made in section~\ref{VI}.

We note that in all the subsequent parts of this paper, we use the formulation ``fire-streaks'' or ``fire-streak approach'' to address our earlier work made for A+A collisions~\cite{1}. This is made to underline the basic similarity of our approach to the original fire-streak concept of~\cite{2}. We note that differences exist on the detailed level, which become clearly apparent from the comparison of our model formulation in section~\ref{II} to the cited original works.



{Finally, we also note 
the
correspondence of our results to the recent works aimed at the explanation of $\Lambda$ and $\overline{\Lambda}$ polarizations observed by the STAR Collaboration in Au+Au collisions~\cite{Starnature}, 
by the initial angular momentum 
generated in 
a
fire-streak-like approach~\cite{Xie2017}. 
How the initial angular momentum is transferred to $\Lambda$/$\bar{\Lambda}$ baryons is still unclear.
It is our hope that the work presented here will bring its modest contribution to 
a better understanding of the applicability of 
the fire-streak-like
approaches to the 
field of high energy reactions.}


\section{The fire-streak fragmentation function}
\label{II}

The model we formulated for ultrarelativistic Pb+Pb collisions, Fig.~\ref{fig1}, assumed the division of the 3D nuclear mass distribution into longitudinal ``bricks'' in the perpendicular plane of the reaction, and the subsequent formation of 
fire-streaks
moving along the collision axis~\cite{1}. 
In the cited reference fire-streaks
of finite transverse size, 1~x~1 fm$^2$, were considered.
Our
fire-streak 
fragmentation function into negative pions 
was parametrized in
the form:

\begin{equation}
  \frac{\mathrm{d}n}{\mathrm{d}y}(y,y_s,E^*_\text{s},m_\text{s}) =
    A \cdot (E^*_\text{s} - m_\text{s})
  \cdot \exp\left(- \frac{[(y - y_s)^2 + \epsilon^2]^\frac{r}{2}}{r \sigma_y^r}\right)  \; \; \; .
  \label{fragmentation}
\end{equation}
The formula~(\ref{fragmentation})
defines the distribution $\frac{\mathrm{d}n}{\mathrm{d}y}$ of 
negative
pions 
created by
the fragmentation of a single fire-streak. 
We named it 
``fire-streak fragmentation function'' 
in order 
to differentiate 
from the 
``standard''
 parton-to-hadron
 fragmentation function (FF)~\cite{xxxxx}. 
In the above,
$y$ is the rapidity of the pion,
$y_s$ is the fire-streak rapidity given by energy-momentum conservation, 
$E^*_\text{s}$
is
its total energy in its own rest frame (or equivalently, its invariant mass, also given by the $E$$-$$\vec{p}$ conservation), and $m_\text{s}$ is the sum of ``cold'' rest masses of the two ``bricks'' forming the fire-streak (given by collision geometry). $\epsilon$ is a small number ensuring the continuity of derivatives ($\epsilon=0.01$ was used in~\cite{1}). Finally, $A$, $\sigma_y$ and $r$ are the only free parameters of the function~(\ref{fragmentation}). 
They appeared common to all the fire-streaks 
in all the collisions,
and independent 
of Pb+Pb
collision centrality\footnote{Deviations from the mean value of $A$ quoted above were smaller or comparable to 
systematical errors of the experimental data~\cite{2.5na49}.
{We note that the numerical values of the parameters discussed in the text apply only to the collision energy $\sqrt{s_{NN}}=17.3$~GeV. The energy dependence of the fire-streak fragmentation function and its parameters,
which emerges from the comparison of Figs~\ref{fig2} and~\ref{fig2-40}
is discussed in detail in~Appendix~B.}}.
%
%
The fit made 
in our
analysis of the NA49 centrality selected Pb+Pb data
{at $\sqrt{s_{NN}}=17.3$~GeV}~\cite{2.5na49}
gave
%
$A=0.05598$, 
$\sigma_y=1.475$, and $r= 2.55$.
%
%
In
this 
analysis,
our 
modelled
pion 
rapidity distribution
in a given centrality selected sample of Pb+Pb collisions of impact parameter $b$ was constructed as the sum of independent fragmentation functions, corresponding to all the constituent fire-streaks:

\begin{equation}
  \frac{\mathrm{d}n}{\mathrm{d}y}(y,b) 
=
  \sum_{(i,j)} \frac{\mathrm{d}n}{\mathrm{d}y}\left(~y,~y_{s_{(i,j)}}(b),~E^*_{\text{s}_{(i,j)}}(b),~m_{\text{s}_{(i,j)}}(b)~\right) \; \; \; \; ,
\label{integrated_dn_dy_approx}
\end{equation} 
where ($i$,$j$) denominate the 
position 
of a given fire-streak
in the transverse ($x,y$) plane of the Pb+Pb collision.
Using formula~(\ref{integrated_dn_dy_approx}), our 
simple
model was able to describe the whole centrality dependence of negative pion $dn/dy$ yields as a function of rapidity, including in particular the narrowing of the rapidity distribution from peripheral to central Pb+Pb collisions as illustrated in Fig.~\ref{fig2}.

Now we proceed to proton-proton collisions, where the total available energy is $\sqrt{s}$. 
We will naively try to apply 
the function~(\ref{fragmentation}) to pion production in the entire p+p system,
 with $E^*_{\text{s}}\rightarrow\sqrt{s}$, $m_\text{s}\rightarrow 2m_\text{p}$. The pion rapidity distribution would then be:

\begin{equation}
  \frac{\mathrm{d}n}{\mathrm{d}y} =
    A   \cdot ({\sqrt{s}} - 2m_\text{p})
   \cdot \exp\left(- \frac{[y^2 + \epsilon^2]^\frac{r}{2}}{r \sigma_y^r}\right) \; \; \; \; ,    
  \label{eq0}
\end{equation}
where $\sqrt{s}=17.27$~GeV as for Pb+Pb collisions, and $m_\text{p}$ is the proton mass.
We note that $y_s=0$ by definition in the p+p c.m. system.
%
Applying $\epsilon=0.01$ and the same parameters 
$A=0.05598$, 
$\sigma_y=1.475$, and $r= 2.55$
which we 
obtained 
from the fit to Pb+Pb collisions~\cite{1}, we 
get
explicitly:

\begin{equation}
  \frac{\mathrm{d}n}{\mathrm{d}y} \equiv
f(y) =
    0.8618
   \cdot \exp\left(- \frac{[y^2 + 0.01^2]^\frac{2.55}{2}}{2.55 \cdot 1.475^{~2.55}}\right) \; \; \; \; \; .
  \label{eq2.3}
\end{equation}
In the following section 
we will 
directly 
compare 
the 
function (\ref{eq2.3}) 
to the experimental 
rapidity distribution in p+p collisions.
We will constantly address $f(y)$ as ``fire-streak fragmentation function'' in the text below, to underline that it was deduced from Pb+Pb reactions as described above.


\section{The negative pion rapidity spectrum}
\label{III}

\begin{figure}[h]
\begin{center}
\hspace*{-1.5cm}\includegraphics[width=10cm,height=8cm]{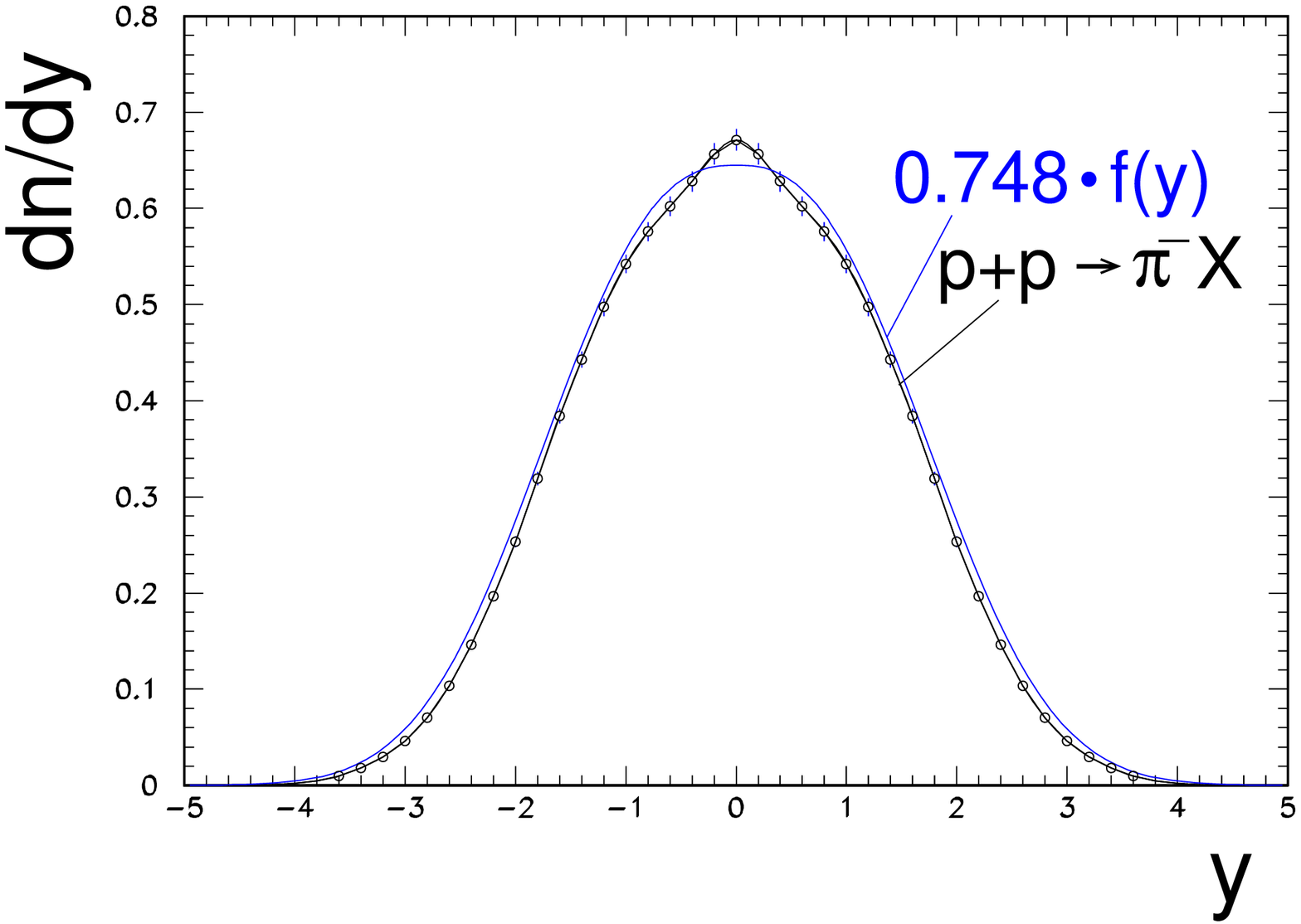}
\vspace*{-0.3cm}
{\caption\small Rapidity distribution of negative pions produced in inclusive inelastic p+p collisions at $\sqrt{s}=17.27$~GeV (experimental data points), compared to our function $f(y)$ from Eq.~(\ref{eq2.3}) multiplied by 
  0.748                
(blue curve).
The 
data points come from~\cite{ppaper} (their
numerical values and errors are taken from~\cite{spshadrons}; only statistical errors are 
shown).
{At negative rapidity
reflected data points are drawn.}\label{fig3}}   
%
%
\end{center}
\end{figure}

The NA49 experiment published rapidity distributions of positively and negatively charged pions in inclusive inelastic p+p collisions at $\sqrt{s}=17.27$~GeV~\cite{ppaper}. 
A comparison of shapes
%
between the experimental 
$p+p\rightarrow\pi^-X$ distribution
and our 
function $f(y)$ defined by Eq.~(\ref{eq2.3}) above 
is 
presented in Fig.~\ref{fig3}. We note that the function $f(y)$ 
multiplied by 
a factor of
0.748 
matches
the experimental data reasonably well. Several facts are noteworthy:
%
%
\begin{itemize}
\item[(1.)]
 It is important to underline that the $p+p\rightarrow\pi^-X$ data in Fig.~\ref{fig3} are compared to the 
{\em single} fire-streak fragmentation function 
function $f(y)$.
This is very different from our study of Pb+Pb collisions made in~\cite{1} and shown in 
{Figs~\ref{fig1}-\ref{fig2-40}.} 
In this latter case our model 
calculation
was always 
the
 {\em sum} of 
fragmentation functions corresponding to the different fire-streaks,
see Eq.~(\ref{integrated_dn_dy_approx}). 
Summing
over many fire-streaks with different values of rapidity $y_S$ affected the width of the overall pion rapidity distribution, which was largest in peripheral and smallest in central Pb+Pb collisions, see {Figs~\ref{fig2}-\ref{fig2-40}}.  
\newpage
\item[(2.)]
 Account taken that 
all the parameters characterizing the function $f(y)$ have been directly inherited from the fit to Pb+Pb collisions,\footnote{We note that the numerical values of $\epsilon$, $\sigma_y$ and $r$ as well as the functional shape
given by Eq.~(\ref{fragmentation}) were
published in~\cite{1} before we started the present analysis.} 
and account taken of the difference between the two analyses stated in (1.),
the 
overall
agreement of the fire-streak functional shape 
with the experimental p+p data is
 in our opinion
{surprisingly good}.
\item[(3.)]
{Notwithstanding the above, 
a deviation of the data points from $f(y)$ can be seen in the central region (most evidently at $y=0$). This goes beyond the 
statistical errors of the data points.  
It is always tempting to discuss such differences in the context of 
systematical errors
of the experimental p+p and Pb+Pb data~\cite{ppaper,2.5na49}, but
it is more natural to explain them by addressing 
the limitations of the procedure for the extraction of the fire-streak fragmentation function which we 
proposed in~\cite{1}. 
Indeed, this function is the result of a non-perturbative process and is only approximated, in an effective way, by our simple formula~(\ref{fragmentation}). Therefore, its extraction from experimental distributions in Pb+Pb collisions, each being a sum of independent fire-streaks according to Eq.~(\ref{integrated_dn_dy_approx}), will smear out all the ``subtleties'' present in the shape of $f(y)$, leaving only its basic smoothened form 
which can be 
described by Eq.~(\ref{fragmentation})
and thus giving the result which we see in Fig.~\ref{fig3}.}
\item[(4.)]
 Finally, a clear discrepancy in the absolute normalization of our function $f(y)$ 
with respect to
the experimental p+p data is evident from~Fig.~\ref{fig3}. This discrepancy, which we attribute to baryon stopping and strangeness enhancement phenomena, will be addressed in 
section~\ref{IV}. 
\end{itemize}

The situation described above, and most of all the somewhat intriguing fact that the experimental $p+p\rightarrow\pi^-X$ distribution can be described,
   or approximated,
by the same shape as that obtained in 
$Pb+Pb\rightarrow\pi^-X$ reactions 
but for the {single fire-streak} (item~(2.)), raises interesting questions. 
Some of these
will be addressed 
in the subsequent parts of this paper. 
%
In the following 
two sections 
we will focus on the difference in 
absolute normalization 
discussed
in item (4.).


\section{Correction for isospin in p+p reactions}
\label{III.5}

\begin{figure}[t]
\begin{center}
\hspace*{-1.5cm}\includegraphics[width=10cm,height=8cm]{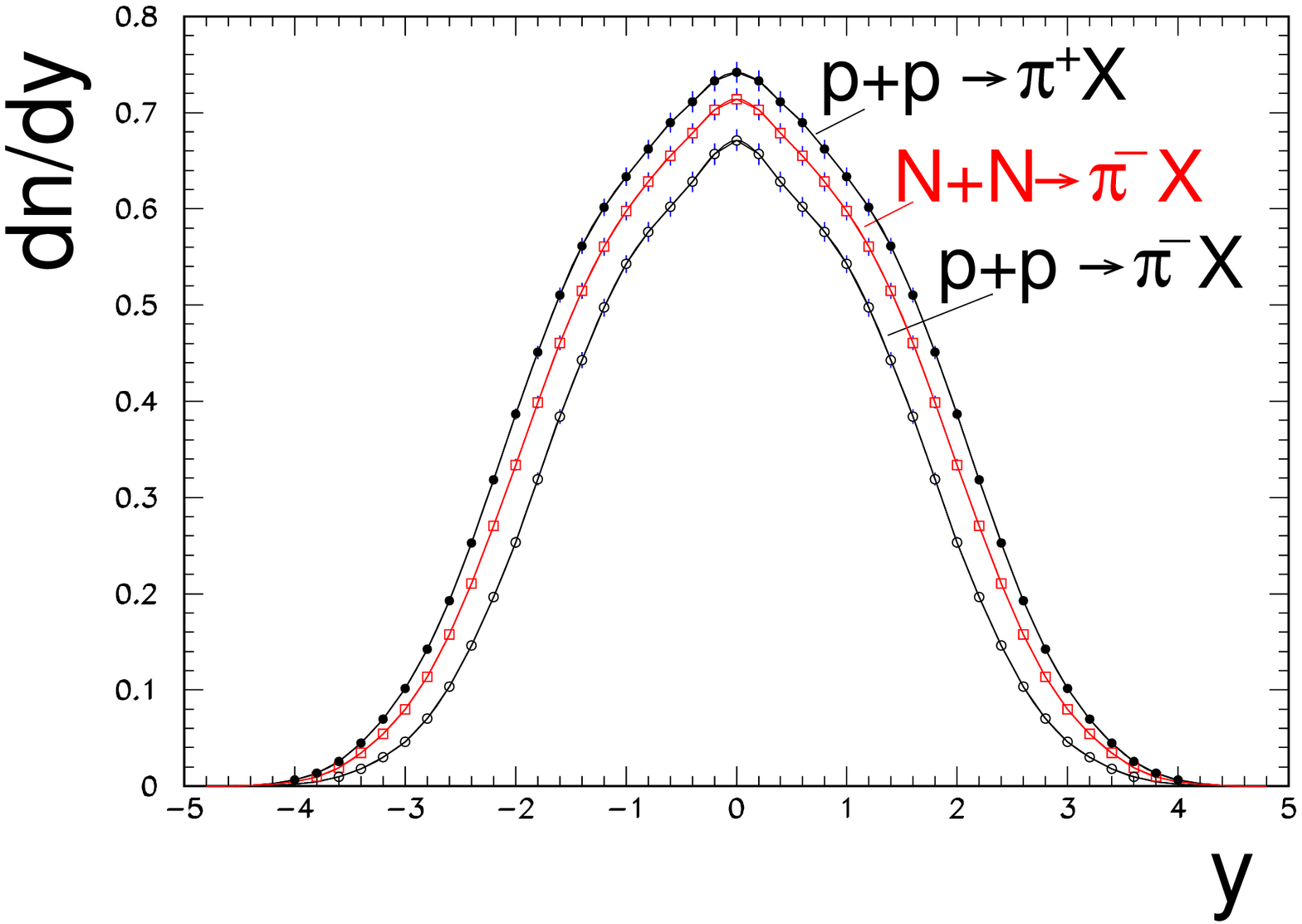}
\vspace*{-0.5cm}
{\caption\small Experimental rapidity
distributions of positive and negative pions produced in inclusive inelastic p+p collisions at $\sqrt{s}=17.27$~GeV (black), together with our isospin-averaged negative pion distribution,
$N$$+$$N$$\rightarrow$$\pi^-$$X$,
 given by Eq.~(\ref{eq2}) (red). The experimental data points come from~\cite{ppaper} (their numerical values and errors are taken from~\cite{spshadrons} and the same relative errors are assumed for the isospin-averaged distribution).
{At negative rapidity
reflected data points are drawn.}\label{fig4}}   
\end{center}
\end{figure}
%
\begin{figure}[t]
\begin{center}
\hspace*{-1.5cm}\includegraphics[width=10cm,height=8cm]{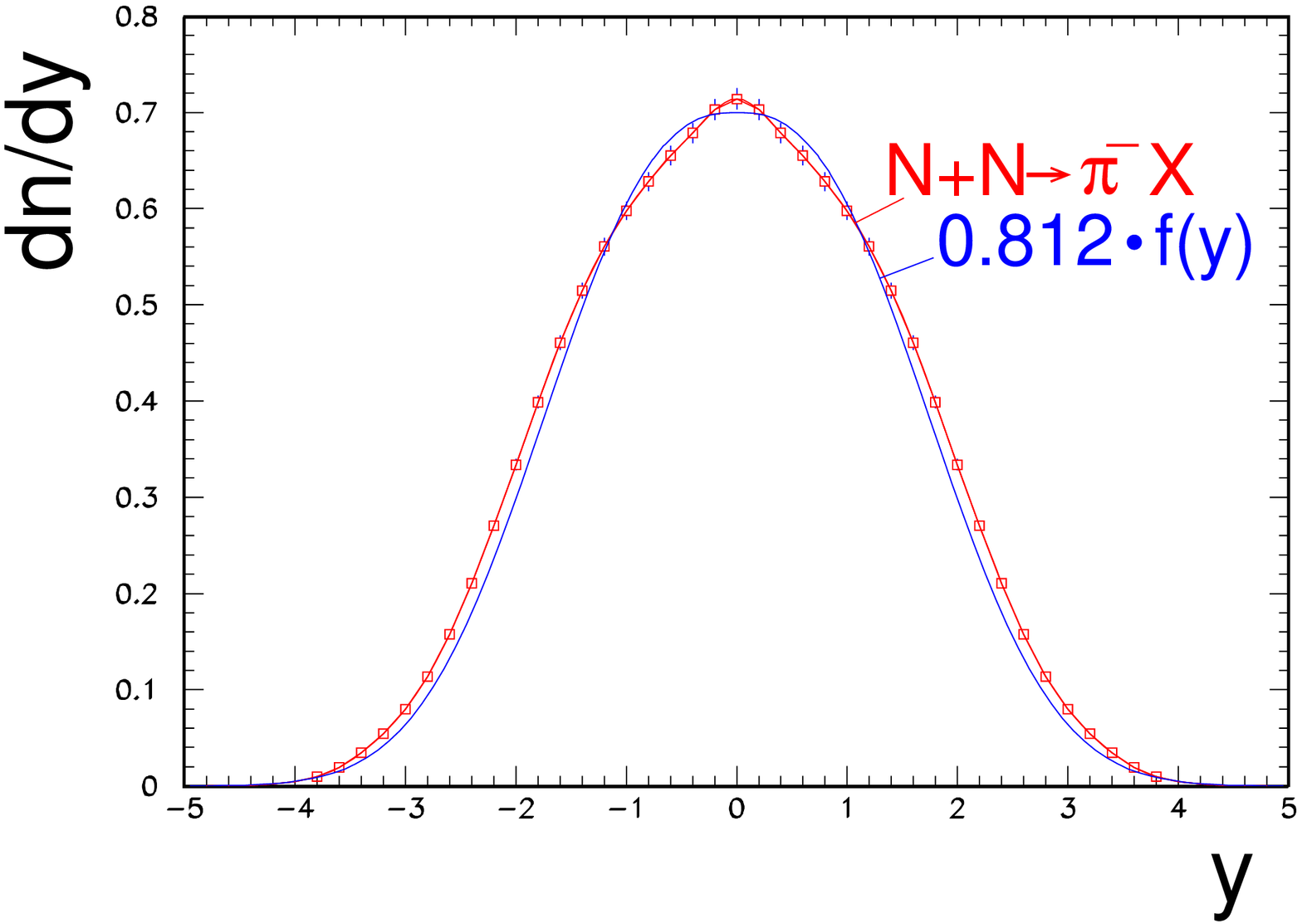}
\vspace*{-0.5cm}
{\caption\small Comparison of negative pion rapidity distribution in inclusive inelastic p+p collisions after correction for isospin effects (red points) to our
single fire-streak 
fragmentation function $f(y)$ from Eq.~(\ref{eq2.3})
(blue curve). The isospin-averaged negative pion distribution $N$$+$$N$$\rightarrow$$\pi^-$$X$ is the same as in Fig.~\ref{fig4}.  
Our function $f(y)$ 
is multiplied by 
0.812.\label{fig5}}  
\end{center}
\end{figure}

As we specified in the precedent section, the single fire-streak 
fragmentation function agrees with the experimental $p+p\rightarrow\pi^-X$ distribution up to a normalization factor of 
0.748. 
Before addressing what we consider as truly dynamical reasons for this difference in normalization, a more ``trivial'' issue is to be addressed.
%
This is 
the difference in the isospin content of the p+p and Pb+Pb systems. As the Pb~($A$=208,~$Z$=82) nucleus consists of $\frac{Z}{A}$=39.4\% protons and $(1-\frac{Z}{A})$=60.6\% neutrons, the proper reference for the $Pb$$+$$Pb$$\rightarrow$${\pi^-}$$X$ spectrum is not the $p$$+$$p$$\rightarrow$$\pi^-$$X$ distribution, but rather that of negative pions obtained from a properly averaged mixture of p+p, n+p, p+n, and n+n collisions. This problem is non-negligible at SPS energies where $\pi^+$ and $\pi^-$ yields in p+p collisions differ quite significantly, as shown in Fig.~\ref{fig4}.

We address this issue by estimating the proper isospin-averaged distribution following the approach proposed in~\cite{x},
invoking isospin symmetry 
in pion production for participating protons and neutrons
$\left(~\frac{\mathrm{d}n}{dy}(n\rightarrow\pi^-) =  
   \frac{\mathrm{d}n}{dy}(p\rightarrow\pi^+)~\right)$.
On that basis the proper ``nucleon+nucleon'' reference for Pb+Pb collisions reads:
%
\begin{equation}
  \frac{\mathrm{d}n}{dy}(N+N\rightarrow\pi^-X)=
  \left(\frac{Z}{A}\right)\cdot
  \frac{\mathrm{d}n}{dy}(p+p\rightarrow\pi^-X)+
  \left(1-\frac{Z}{A}\right)\cdot
  \frac{\mathrm{d}n}{dy}(p+p\rightarrow\pi^+X) \; \; \; .
  \label{eq2}
\end{equation}
The distribution~(\ref{eq2}) is presented in Fig.~\ref{fig4}. 
In Fig.~\ref{fig5}, its
shape is compared to our 
function $f(y)$ given by Eq.~(\ref{eq2.3}). 
We consider that after the correction for isospin differences, the agreement of the $N+N\rightarrow\pi^-X$ distribution
with 
$f(y)$ 
- the latter being inherited from our description of 
the
Pb+Pb reactions as 
explained
in section~\ref{II} -
is
invariably
good. 
The normalization factor increases from
%
0.748 to 0.812.

In the next section we will attempt to understand 
this factor. 

\vspace*{-1cm}
\section{The absolutely normalized pion yield in p+p 
collisions}
\label{IV}

In the following we will use energy conservation 
to estimate whether the agreement apparent in the 
comparison of the distribution shapes, in 
Figs~\ref{fig3} and~\ref{fig5}, can be reconciled with the fact that our 
function $f(y)$, 
derived from 
Pb+Pb reactions,
brings a total pion yield which is
evidently 
higher than what is measured in p+p collisions.
This difference 
in total pion yield
is
quantified
(after correction for isospin effects)
by the 
normalization factor 
    0.812
addressed above.
%
We consider it
conceivable that specific dynamical mechanisms, similar in p+p and Pb+Pb collisions (dressing up of quarks into hadrons to quote the first example) would lead to a similar {\em shape} of longitudinal distributions of final state particles, while the absolutely normalized final production {\em yields} would be 
significantly 
different.
%
Therefore we will consider the
differences in the overall energy balance between 
nucleon-nucleon
and nucleus-nucleus reactions, that is, the different repartition of collision energy into the various types of final state particles.



We see two main, experimentally well established, phenomena which
modify
this energy balance. These are:

\begin{itemize}
\item[(1.)]
Baryon stopping~\cite{busza84}, i.e. the change in baryon inelasticity between p+p and Pb+Pb collisions;
\item[(2.)]
Strangeness enhancement, that is the enhanced production of strange over non-strange particle production, since a long time interpreted as connected to quark gluon plasma formation in heavy ion reactions~\cite{rafelski}.
\end{itemize}

The influence of these two phenomena on the overall energy 
repartition in p+p and Pb+Pb reactions
will be estimated below.
We underline that the aim of this section is to provide both estimates in a maximally model-independent way. For this reason, we decide not to use 
any particular model for baryon stopping or strangeness enhancement 
which would need to be validated against experimental data. Instead, we decide to study these issues using experimental data {directly}, whenever available. As this will become apparent below, the fact that such a 
study can be made with a reasonable precision speaks very well for the completeness of experimental information at SPS energy.

Consequently,
the work described in sections~\ref{IV.1} and~\ref{IV.2} is to be understood as an attempt at a fair comparison of the results of our work on Pb+Pb collisions with experimental p+p data as we said in section~\ref{intro}, and not as an extension of the fire-streak model of Pb+Pb collisions to p+p reactions. 
\subsection{Baryon stopping}
\label{IV.1}

Uniquely for clarity and conciseness, 
the discussion made below will {\em implicitly} include the correction for isospin differences between p+p and Pb+Pb reactions addressed in section III above. Thus we will assume that 
formula~(\ref{eq2}) correctly describes the mixture of nucleon+nucleon (p+p, n+p, p+n and n+n) collisions representative for Pb+Pb reactions, and concisely write
\begin{equation}
  \frac{\mathrm{d}n}{dy}(p+p)
~~~~~~\text{instead~of}~~~~~~
  \frac{\mathrm{d}n}{dy}(N+N\rightarrow\pi^-X)
  \label{eqconcise}
\end{equation}
for the representative, isospin corrected distribution from 
Eq.~(\ref{eq2}). 
Consequently whenever we refer to ``p+p'' (or ``$pp$'') reactions, the representative set of nucleon+nucleon collisions will be meant. 
Also, we will
neglect the small difference between 
proton and neutron masses.
Finally, for simplicity, we will apply the convention $\sqrt{s_{NN}}\equiv\sqrt{s}$ independently on the considered reaction type.

Let us now consider the agreement of rapidity 
distribution
shapes shown in Fig.~\ref{fig5}, together with our 
formulae~(\ref{eq0}) and~(\ref{eq2.3}).
Approximately, we can quantify this 
agreement 
as follows:
\begin{equation}
  \frac{\mathrm{d}n}{\mathrm{d}y} (p+p)
=
    A_{pp}   \cdot (\sqrt{s}   - 2m_\text{p})
   \cdot \exp\left(- \frac{[y^2 + \epsilon_{AA}^2]^\frac{r_{AA}}{2}}{r_{AA} \cdot \sigma_{y_{AA}}^{r_{AA}}}\right) \; \; \; \; ,
  \label{eq4}
\end{equation}
%
%
where we put explicitly 
$\epsilon_{AA}=0.01$, 
$\sigma_{y_{AA}}=1.475$, and $r_{AA}= 2.55$ 
 to underline that these parameters are obtained from $AA$ (Pb+Pb) reactions with no further tuning to p+p collisions. On the other hand the normalization parameter $A_{pp}$ is specific to the p+p reactions. We know from Fig.~\ref{fig5} that
\begin{equation}
     A_{pp}=0.812 \cdot A_{_{AA}} \approx 0.8 \cdot A_{_{AA}}  \; \; \; \; \; , 
\label{eq4.5}
\end{equation}
%
     where $A_{_{AA}}=0.05598$ was obtained from 
experimental data on
Pb+Pb collisions as specified in section~\ref{II}.

Let us now consider a central Pb+Pb collision at impact parameter $b\approx 0$. As it can be immediately seen from the energy-momentum conservation considerations 
made in
our earlier work~\cite{1}, 
our
model predicts, for such a collision, the formation of fire 
streaks -
all of them build of symmetric ``bricks'' of equal mass and being
at rest in the collision c.m. system ($y_s\approx 0$). For any given fire-streak made of two bricks of 
equal
mass $M$ 
the outcoming $\pi^-$ distribution will be, from Eq.~(\ref{fragmentation}):
\begin{equation}
\begin{aligned}
  \frac{\mathrm{d}n}{dy}(A+A\rightarrow\pi^-X) 
 \equiv
  \frac{\mathrm{d}n}{\mathrm{d}y} (A+A) 
&  =
  A_{_{AA}} \cdot (E^*_\text{s} - m_\text{s})
  \cdot \exp\left(- \frac{[(y - y_s)^2 + \epsilon_{AA}^2]^\frac{r_{AA}}{2}}{r_{AA} \cdot \sigma_{y_{AA}}^{r_{AA}}}\right) \\
&  =
A_{_{AA}} \cdot (E^*_\text{s} - m_\text{s})
  \cdot \exp\left(- \frac{[y^2 + \epsilon_{AA}^2]^\frac{r_{AA}}{2}}{r_{AA} \cdot \sigma_{y_{AA}}^{r_{AA}}}\right) \\
&  =
A_{_{AA}} \cdot (M/m_\text{p}\cdot\sqrt{s} - 2M) \cdot F_{AA}(y) \\
&  =
A_{_{AA}} \cdot B_M \cdot (\sqrt{s} -2m_\text{p}) \cdot F_{AA}(y) \; \; \; \; \; ,
%
%
\label{eq5}
\end{aligned}
\end{equation}
where we introduced the shape factor 
$F_{AA}(y)=\exp\left( - [       y^2 + \epsilon_{AA}^2]^{r_{AA}/2} / (r_{AA} \cdot \sigma_{y_{AA}}^{r_{AA}}) ~ \right)$. 
We note that 
$B_M=M/m_\text{p}$ is the baryon number of each ``brick'' (equivalent to the number of participating nucleons per fm$^2$ in the plane perpendicular to the collision axis). For p+p collisions we rewrite Eq.~(\ref{eq4}) 
in the same form as~(\ref{eq5}):
%
\begin{equation}
  \frac{\mathrm{d}n}{dy}(p+p)=
 A_{pp} \cdot B_M \cdot (\sqrt{s} - 2m_\text{p}) \cdot F_{AA}(y)   \; \; \; \; \; ,
\label{eq6}
\end{equation}
where $B_M=M/m_\text{p}=1$ for p+p reactions.

Let us now relate the energy available for particle production per incoming nucleon pair, 
to the outcoming baryon inelasticity $K$~\cite{blume2007} in the final state of the collision:
\begin{equation}
K=\frac{2\cdot E_{inel}}{\sqrt{s}-2m_\text{p}}   \;  \; \; , 
\label{k}
\end{equation}
where $E_{inel}$ is the total energy lost by the incoming baryon
 which 
remains available for particle production. Let us first assume that the available energy repartition between the different 
types of produced particles
(that is, $\pi^{+}$, $\pi^{-}$, $\pi^0$, kaons, etc) 
remains the same between (isospin-corrected) p+p and Pb+Pb collisions\footnote{This assumption will be re-discussed in sections~\ref{IV.2} and~\ref{IV.3}.}. 
Then we have for the rapidity distribution of negative pions,
respectively from Eqs.~(\ref{eq6}) and~(\ref{eq5}):

\begin{equation}
dn/dy(p+p) = B_M \cdot \tilde{A} \cdot 2E_{inel} \cdot F_{AA}(y) \; \; \; \; \; , 
\label{eq7}
\end{equation}
\begin{equation}
dn/dy(A+A) = B_M \cdot \tilde{A} \cdot 2E_{inel} \cdot F_{AA}(y) \; \; \; \; \; , 
\label{eq8}
\end{equation}
where $\tilde{A}$ in now assumed to be a {\em constant} factor.
From 
(\ref{eq5}), (\ref{eq6}), (\ref{eq7}), and (\ref{eq8}) 
we have:
%
\begin{equation}
A_{pp}=\tilde{A}\cdot K_{pp}  \; \; \; \; \; ,
\label{eq9}
\end{equation}
\begin{equation}
A_{AA}=\tilde{A}\cdot K_{AA}  \; \; \; \; \; .
\label{eq10}
\end{equation}
Thus under the assumption
made
above, the difference in normalization of 
pion
rapidity distributions in proton-proton 
reactions 
and 
in
a single fire-streak from
the
Pb+Pb collisions 
(Figs~\ref{fig3} and~\ref{fig5})
would come from differences in final state baryon inelasticity. 

\begin{table}
\begin{tabular}{|c|c|c|c|}
  \hline
Reaction & ~~~$p+p\rightarrow(p-\bar{p})X$~~~  
         & ~~~$p+p\rightarrow(B-\bar{B})X$~~~   
         & ~~~$Pb+Pb\rightarrow(p-\bar{p})X$~~~  \\ \hline
Ref.     & ~\cite{pprot,spshadrons} &  ~\cite{pprot,spshadrons}&  ~\cite{blume2007}         \\ \hline
$K$        &   0.522         &   0.547         &   0.78           \\ \hline
        \multicolumn{4}{|c|}{ ~~~~~~~~~~~~~~~~~~~~~~~~~~~~~~~~~~~~~~~~~~~~
~~~~~~~~~~~~~~~~~~~~~
 ratio $K_{pp}/K_{{AA}}$ = 0.70  }        \\ \hline
\end{tabular}
\caption{Compilation of our knowledge on baryon inelasticity 
in p+p and central Pb+Pb collisions at $\sqrt{s}_{NN}=17.27$~GeV.
The value in the middle column includes both net protons and net neutrons as described in the text.}
\label{XXX}
\end{table}

Here 
a lot of
information is available at SPS energies.
For {\em proton-proton reactions}, the common knowledge in the community is that the proton looses about half of its energy in the collisions~\cite{strob}, which gives $K_{pp}\approx 0.5$. It is to be noted that the $p+p\rightarrow pX$ distribution, best known experimentally, may be subject to isospin effects 
if compared to Pb+Pb reactions where more neutrons participate than protons. Both statements can at present be verified with 
experimental data from 
the
NA49~\cite{pprot} and NA61/SHINE~\cite{NA61_pp} collaborations. In particular, the NA49 reference~\cite{pprot} includes not only 
precise, double differential in ($x_F$,$p_T$), very wide acceptance proton and antiproton data, but also the 
neutron $x_F$ distribution at $\sqrt{s}=17.27$~GeV. The cited paper includes also a precise numerical interpolation of the $p$ and $\bar{p}$ data~\cite{spshadrons} which can 
be used to obtain
a model-independent evaluation of net proton inelasticity. 
We underline 
again
the superiority of using such 
a
wide acceptance 
interpolation of
experimental data rather than relying on a particular model-dependent event generator.
We performed this evaluation 
and obtained $K=0.522$ as shown in Table~\ref{XXX}. This was made by calculating numerically the average net proton energy in an inclusive inelastic p+p event and consequently obtaining $E_{inel}$ in Eq.~(\ref{k}):
\begin{equation}
E_{inel}=\frac{\sqrt{s}}{2}-\langle E_\text{net~proton}\rangle \; \; \; \; \; ;
\text{~~~~~~~~~~~with}
\label{einel}
\end{equation}
\begin{equation}
\langle E_\text{net~proton}\rangle
 ~=~ 
\frac{\int_{0}^{1}\int_{0}^{p_T(\mathrm{max})}E(x_F,p_T)\cdot \left(\frac{d^2\sigma}{dx_Fdp_T}\right)_\text{net~proton} {dp_T~dx_F}}{\int_{0}^{1}\int_{0}^{p_T(\mathrm{max})}\left(\frac{d^2\sigma}{dx_Fdp_T}\right)_\text{net~proton} {dp_T~dx_F}}
\; \; \; \; \; ,
\label{e_av_netp}
\end{equation} 
where $E(x_F,p_T)$ is the net proton energy 
given by its $x_F$ and $p_T$, 
and the net proton density is obtained by the subtraction of the quoted interpolated proton and antiproton distributions:
\begin{equation}
\left(\frac{d^2\sigma}{dx_Fdp_T}\right)_\text{net~proton} 
 ~=~ 
\left(\frac{d^2\sigma}{dx_Fdp_T}\right)_{p} 
 ~-~ 
\left(\frac{d^2\sigma}{dx_Fdp_T}\right)_{\overline{p}}\; \; \; \; \; .
\label{d2ndxfdpt_netp}
\end{equation} 
%
We note that the numerical integration in Eq.~(\ref{e_av_netp}) above was performed assuming $p_T(\mathrm{max})=2$~GeV/c, over a grid of 1000~x~1000 sampling points.

Subsequently, on the basis of the same data interpolation as well as of the published experimental neutron $x_F$ distribution, we estimated the (net proton)+(net neutron) spectrum assuming that neutrons have the same 
shape of the
$p_T$ distribution as protons at a given $x_F$, an assumption that should have 
only 
a small influence on the final result. Following the considerations about antineutrons made in~\cite{pprot}, we subtracted 1.66~times~(see~\cite{pprot}) the antiproton distribution in order to obtain the net neutron spectrum. We applied formulae strictly similar to (\ref{einel})-(\ref{d2ndxfdpt_netp}), as well as the same integration sampling grid and limits.
The final result 
for net baryons (protons+neutrons) in the final state of the p+p collision 
is $K_{pp}=0.547$, as 
shown in Table~\ref{XXX}. We note that this result is already free from isospin effects as it contains both isospin partners. We neglect the contribution of other baryons like $\Lambda$ due to their small cross-section.

For {\em central Pb+Pb collisions}, we expect that the lower acceptance coverage of existing experimental distributions may induce 
a stronger model dependence for the estimate on $K_{AA}$. On the other hand, the net proton distribution in Pb+Pb collisions should be weakly 
affected by
isospin effects due to the mixed isospin content of the lead nucleus. All in all, we consider the estimate provided by C.~Blume~\cite{blume2007}, where the contribution of unmeasured baryons was estimated from the statistical hadron gas model~\cite{4becattini} as secure enough for our study. The latter gives $K_{AA}\approx 0.78$ at top SPS energy.

From the above, we estimate from~(\ref{eq9}) and (\ref{eq10}):

\begin{equation}
A_{pp} / A_{AA} = K_{pp} / K_{AA} =       0.547 / 0.78 \approx 0.70 \; \; \; \; \; \; . 
\label{eq0.68}
\end{equation}
This is to be compared 
to
    $A_{pp}/A_{AA}=0.812$ 
established from Fig.~\ref{fig5} in section~\ref{III}. Thus we see that energy conservation-related considerations connected to changes in baryon inelasticity can explain a part of the normalization difference between the experimental pion 
rapidity
spectrum 
in inelastic p+p collisions, and that obtained from a single fire-streak in Pb+Pb reactions. However, our result
overpredicts the difference which we saw in Fig.~\ref{fig5}: the fire-streak fragmentation function matches the shape of the experimental $p+p\rightarrow\pi^-X$ spectrum, but the difference in the absolute normalization of the two distributions is {\em smaller}
than 
what is
expected solely from differences in inelasticity.

\subsection{Strangeness enhancement}
\label{IV.2}

It is very well known that production of strange particles (mostly $K$ mesons~\cite{aduszkiewicz}, but also strange baryons~\cite{na57}) is significantly enhanced in Pb+Pb with respect to p+p collisions. 
%
In the following we refrain from discussing the dynamical origin of strangeness enhancement which has been done before 
in 
very 
well known papers~\cite{rafelski, smes}. We focus on the energy balance between strange and non-strange particle production. For simplicity we limit ourselves to pions and kaons which dominate the yields of  produced particles. 
The changes in baryon inelasticity must 
also be taken into account.

Table~\ref{XXXX} displays our compilation of kaon and pion yields in central Pb+Pb as well as p+p collisions, taken together with mean pion and kaon energies in inelastic p+p events at 
the
top SPS energy. The latter should be commented upon. The presented estimates are in our view completely model-independent as they are uniquely based on very detailed and wide acceptance two-dimensional ($x_F$,$p_T$) distributions from the NA49 experiment~\cite{ppaper,kaonpaper}. Precise numerical interpolations of these distributions have been included therein and remain available in~\cite{spshadrons}.
Our estimates for mean energies are directly, numerically computed from these interpolated experimental distributions. 
For this purpose we use a formula similar to~(\ref{e_av_netp}):
\begin{equation}
\langle E_{i}\rangle
 ~=~ 
\frac{\int_{0}^{1}\int_{0}^{p_T(\mathrm{max})}E_{i}(x_F,p_T)\cdot \left(\frac{d^2\sigma}{dx_Fdp
_T}\right)_{i} {~~dp_T~dx_F}}{\int_{0}^{1}\int_{0}^{p_T(\mathrm{max})}\left(\frac{d^2\sigma}{dx_Fdp_T}\right)_{i} {~~dp_T~dx_F}}
\; \; \; \; \; ,
\label{e_av_i}
\end{equation} 
where $i$ denotes the particle type $(i=\pi^+,\pi^-,K^+,K^-)$, for which the production cross section $\left(\frac{d^2\sigma}{dx_Fdp_T}\right)_{i}$ has been measured and numerically interpolated over a very large phase space in~\cite{ppaper,kaonpaper}. $E_{i}(x_F,p_T)$ denotes the particle's energy at a given $(x_F,p_T)$ which is uniquely defined by its mass $(m_i=m_\pi~\text{or}~m_K)$. Thanks to the symmetry of the p+p collision we can limit the integration to positive $x_F$ 
only. 
We apply $p_T(\mathrm{max})=2$~GeV/c, and a grid of 1000~x~1000 sampling points.
Here we wish to emphasize again the value of these precisely interpolated data provided by~\cite{pprot,ppaper,kaonpaper}, as well as the advantage of our model-independent approach with respect to both 
model simulations as well as simple analytical parametrizations of experimental data.

\begin{table}
\begin{tabular}{|c|c|c|c|c|}
  \hline
  \multirow{2}{*}{Reaction} & 
     \multicolumn{4}{|c|}{total average yield per event}\\ \cline{2-5}
                          & ~~~~~$\pi^+$~~~~~ & ~~~~~$\pi^-$~~~~~~ & ~~~~~$K^+$~~~~~ & ~~~~~$K^-$~~~~~ \\ \hline
  central Pb+Pb,          &    560  &  602  &   97.8 &  54.0 \\
  $\sqrt{s_{NN}}=17.27$~GeV & ~\cite{tof} & ~\cite{2.5na49}&  ~\cite{2.5na49}&  ~\cite{2.5na49} \\ \hline
  \multirow{3}{*}{inelastic p+p,}&   3.018 &  2.360  & 0.2267& 0.1303 \\
  \multirow{3}{*}{$\sqrt{s_{NN}}=17.27$~GeV} & ~\cite{ppaper} &~\cite{ppaper} &~\cite{kaonpaper}& ~\cite{kaonpaper} \\ \cline{2-5}
   & \multicolumn{4}{|c|}{average energy per particle [MeV]}\\ \cline{2-5}
   &   905   &  781  &   1388  & 1107   \\ \hline
\end{tabular}
\caption{Charged pion and kaon yields in central Pb+Pb and inelastic p+p collisions at top SPS energy, put together with our estimates of mean pion and kaon energy in inelastic p+p collisions obtained numerically from interpolated experimental data as discussed in the text. The quoted values are taken from the references cited in the table.}
\label{XXXX}
\end{table}

In the following we 
will
assume 

\begin{equation}
\begin{aligned}
 \pi^0 & \approx \frac{\pi^++\pi^-}{2} \; \; \; , \\ 
K^0  + & \overline{K}^0  \approx K^++K^- \; \; \;  
\end{aligned}
\label{XA}
\end{equation}
for these particles' kinematical spectra and average yields; we consider these rough assumptions 
to be 
good enough for our present 
evaluation.
On that basis,
from Table~\ref{XXXX} we obtain the average total energy 
which
an inelastic p+p collision 
will spend 
on pion, 
$K^+$, $K^-$ and ($K^0+\overline{K}^0$) production. These we denote as $E(pp\rightarrow\pi)$, where $\pi\equiv (\pi^++\pi^-+\pi^0)$, 
and then respectively
$E(pp\rightarrow K^{+})$, 
$E(pp\rightarrow K^{-})$, 
and $E(pp\rightarrow K^{0\overline{0}})$
where $K^{0\overline{0}}\equiv (K^0+\overline{K}^0)$. 

\begin{equation}
\begin{aligned}
& E(pp\rightarrow\pi)             = 3/2 \cdot (3.018 \cdot 905 + 2.360 \cdot 781) = 
6862~\mathrm{MeV} \; \; \; , \\ 
& E(pp\rightarrow K^{+})          =  0.2267 \cdot 1388 = 315~\mathrm{MeV} \; \; \; , \\ 
& E(pp\rightarrow K^{-})  =  0.1303 \cdot 1107 = 144~\mathrm{MeV} \; \; \; ,\\
& E(pp\rightarrow K^{0\overline{0}})  =~~ 315 + 144 ~~~= 459~\mathrm{MeV} \; \; \; .
\end{aligned}
\label{XPP}
\end{equation}

As we consider the above values to be useful for future studies, we include them in Table~\ref{X5} together with values
of kaon/pion ratios in p+p and central Pb+Pb reactions extracted from Table~\ref{XXXX} on the basis of assumptions~(\ref{XA}). 
In addition, 
we calculate the ratios of energy spent on kaons ($K^+$, $K^-$ and 
$K^0$$+$$\overline{K}^0$)
relative to that spent on pions ($\pi^++\pi^-+\pi^0$) in 
p+p reactions and in central Pb+Pb collisions. These are respectively:

\begin{table}
\begin{tabular}{|c|c|c|c|c|}
  \hline
  \multirow{2}{*}{Reaction} & 
     \multicolumn{4}{|c|}{kaon/pion ratios}\\ \cline{2-5}
                          & ~~~~~$K^+/\pi$~~~~~ & ~~~~~$K^-/\pi$~~~~~~ & 
                     \multicolumn{2}{|c|}{$(K^0+\overline{K}^0)/\pi$} \\ \hline
  central Pb+Pb,           & 0.0561  & 0.0310 &  \multicolumn{2}{|c|}{0.0871} \\
  $\sqrt{s_{NN}}=17.27$~GeV &         &        &  \multicolumn{2}{|c|}{      } \\ \hline
  \multirow{3}{*}{inelastic p+p,}&  0.0281  & 0.0162 &  \multicolumn{2}{|c|}{0.0443} \\ \cline{2-5}
  \multirow{3}{*}{$\sqrt{s_{NN}}=17.27$~GeV} &  \multicolumn{4}{|c|}{average energy per particle type [MeV]}\\ 
   &  $E(pp\rightarrow\pi)$   &   $E(pp\rightarrow K^{+})$ 
  &   $E(pp\rightarrow K^{-})$  & $E(pp\rightarrow K^{0\overline{0}})$   \\ \cline{2-5}
   &   6862  &   315 &      144 &      459 \\ \hline
\end{tabular}
\caption{Kaon over pion ratios in central Pb+Pb and 
inclusive inelastic
p+p reactions, and
average energies spent on pion and kaon production in a single inelastic p+p event.
By pion ($\pi$) the summed $\pi$ mesons $(\pi^++\pi^-+\pi^0)$ are 
meant.}
\label{X5}
\end{table}

\begin{equation}
\begin{aligned}
  R_\text{energy}(pp\rightarrow& K^{+}/\pi) 
= \frac{E(pp\rightarrow K^{+})}{E(pp\rightarrow\pi)} 
= \frac{315~\mathrm{MeV}}{6862~\mathrm{MeV}}
= 0.04590 \; \; \;  ,\\
\end{aligned}
\label{XPPB}
\end{equation}
\begin{equation}
\begin{aligned}
  R_\text{energy}(pp\rightarrow& K^{-}/\pi) 
= \frac{E(pp\rightarrow K^{-})}{E(pp\rightarrow\pi)} 
= \frac{144~\mathrm{MeV}}{6862~\mathrm{MeV}}
= 0.02099 \; \; \;  ,\\
\end{aligned}
\label{XPPC}
\end{equation}
\begin{equation}
\begin{aligned}
  R_\text{energy}(pp\rightarrow& K^{0\overline{0}}/\pi) 
= \frac{E(pp\rightarrow K^{0\overline{0}})}{E(pp\rightarrow\pi)} 
= \frac{459~\mathrm{MeV}}{6862~\mathrm{MeV}}
= 0.06689 \; \; \;  ,\\
\end{aligned}
\label{XPPCC}
\end{equation}
\begin{equation}
\begin{aligned}
  R_\text{energy}(pp\rightarrow&\text{all~kaons}/\pi) = 
0.04590 +
0.02099 +
0.06689
= 0.13378  \; \; \;  ,\\
%
\end{aligned}
\label{XPPD}
\end{equation}
%

%
\begin{equation}
\begin{aligned}
& R_\text{energy}(PbPb\rightarrow K^{+}/\pi) 
= \frac{\frac{K^+}{\pi}(PbPb)}{\frac{K^+}{\pi}(pp)~~~}
\cdot R_\text{energy}(pp\rightarrow K^{+}/\pi) 
= 0.09164  \; \; \; \; \; \; \;   ,\\
\end{aligned}
\label{XB}
\end{equation}
\begin{equation}
\begin{aligned}
& R_\text{energy}(PbPb\rightarrow K^{-}/\pi) 
= \frac{\frac{K^-}{\pi}(PbPb)}{\frac{K^-}{\pi}(pp)~~~}
\cdot R_\text{energy}(pp\rightarrow K^{-}/\pi) 
= 0.04017  \; \; \; \; \; \; \;   ,\\
\end{aligned}
\label{XC}
\end{equation}
\begin{equation}
\begin{aligned}
& R_\text{energy}(PbPb\rightarrow K^{0\overline{0}}/\pi) 
= \frac{\frac{K^0+\overline{K}^0}{\pi}(PbPb)}{\frac{K^0+\overline{K}^0}{\pi}(pp)~~~~~}
\cdot R_\text{energy}(pp\rightarrow K^{0\overline{0}}/\pi) 
= 0.13152  \; \;     ,\\
\end{aligned}
\label{XCC}
\end{equation}
\begin{equation}
\begin{aligned}
&  R_\text{energy}(PbPb\rightarrow\text{all~kaons}/\pi) =
  0.09164 +
  0.04017 +
  0.13152
= 0.26333  \; \; \;  \; \; \; \; .\\
%
\end{aligned}
\label{XD}
\end{equation}
\vspace*{0.1cm}

We note that in Eqs.~(\ref{XB})-(\ref{XCC}) above, we make the important assumption that the ratio of average energy of one kaon over that of one pion remains constant between inelastic p+p and central Pb+Pb collisions. This assumption, which we consider good enough for our present evaluation, calls for an experimental verification. However, we note 
that as this requires a precise knowledge of $d^2n/dydp_T(y,p_T)$ distributions over a very wide range of both $y$ and $p_T$, a model-independent evaluation of these quantities 
in Pb+Pb collisions
seems 
difficult 
on the level of accuracy attainable for 
the
p+p data,
summarized by Eq.~(\ref{XPP}). 
Under this assumption we see that the kaon contribution to the overall energy balance,
evaluated with respect to that of
pion emission, changes by 
a
factor of about two: from $13\%$ in
inelastic
p+p 
to $26\%$ in central Pb+Pb reactions.

\subsection{Energy balance in particle emission}
\label{IV.3}


We will now estimate the basic balance of energy in the 
emission
of 
strange and non-strange particles in the final state of p+p and Pb+Pb reactions. This we will do 
to investigate 
whether it can explain the differences in the absolute pion yield
between the experimental spectrum in p+p collisions and the 
fire-streak fragmentation function which we 
obtained
from 
the Pb+Pb data
(sections~\ref{III} and~\ref{III.5}). In p+p collisions, the inelastic energy (difference between baryon energy in the initial and the final state) writes:



\begin{equation}
E_{inel} \approx \mathrm{(pion~energy)}+\mathrm{(kaon~energy)} \; \; \; ,\\
\label{YPP}
\end{equation}
where by ``$\approx$'' we 
mean that we neglect particles not considered in our discussion, i.e., mainly 
baryon and anti-baryon pairs as well as strange baryons (mainly $\Lambda$). We justify this assumption by the approximate character of our evaluation. 
Furthermore,
%
we state that our estimated overall energy balance in inelastic p+p collisions holds within 3.7\% 
even when we omit the above particles. The corresponding estimate, and a demonstration of even better consistency after the inclusion of non-strange baryon-antibaryon pairs,
are presented in Appendix~A.

Account taken of the quantitative relations described in 
sections~\ref{IV.1} and~\ref{IV.2}
(formula~(\ref{XPPD})),
 Eq.~(\ref{YPP}) writes:
%
\begin{equation}
E_{inel}(K=0.547) \approx \mathrm{(pion~energy)}\cdot(1 + 0.13378) \; \; \; ,\\
\label{YPPK}
\end{equation}
where $K$ is the baryon inelasticity obtained in section~\ref{IV.1}.
In central Pb+Pb collisions, 
from formula~(\ref{XD})
the corresponding energy balance writes:

\begin{equation}
E_{inel}(K=0.78) \approx \mathrm{(pion~energy)}\cdot(1 + 0.26333) \; \; \; ,\\
\label{YPBPBK}
\end{equation}
where the left term is given by the change in baryon inelasticity and the right term by 
the
strangeness enhancement.

Thus the inelastic energy ``lost'' by one 
incoming
baryon and spent on pion production changes from p+p to central Pb+Pb collisions. It increases by the enhancement of baryon inelasticity but then decreases by the different sharing between pions and particles containing strange quarks. The overall change of energy spent on pion production can thus be described as:

\begin{equation}
\frac{\text{Energy~spent~on~pions~in~Pb+Pb}}{\text{Energy~spent~on~pions~in~p+p}}
=
 \frac{0.78 / (1 + 0.26333)}{0.547 / (1 + 0.13378)}
=
 1.280
=
 \frac{1}{0.781} \approx \frac{1}{0.70}\cdot 0.9 \; ,
\label{XUXU}
\end{equation}
%
%
where the last transformation states explicitly the terms induced by the change in inelasticity (section~\ref{IV.1}) and by 
the
strangeness enhancement (section~\ref{IV.2}).

\subsection{Normalization of pion emission in p+p and Pb+Pb collisions}
\label{IV.4}

Now
let us 
calculate the relative normalization of the pion rapidity distribution in p+p collisions, with respect to that of the fire-streak fragmentation function obtained from the Pb+Pb data (Fig.~\ref{fig5}). 
%
Eqs.~(\ref{YPPK}),~(\ref{YPBPBK}) 
quantify the fact that 
the amount of inelastic energy available for particle production,
and
its sharing
between the emission of particles containing and 
not containing strange quarks, 
are both different in
p+p and Pb+Pb
collisions.
Consequently,
Eqs.~(\ref{eq5})-(\ref{eq6}),~(\ref{eq7})-(\ref{eq8}), and (\ref{eq9})-(\ref{eq10}) get 
rewritten in a new form which explicitly takes 
both issues
into account. This gives respectively the formulae~(\ref{eq15})-(\ref{eq16}), (\ref{eq17})-(\ref{eq18}), and~(\ref{eq19})-(\ref{eq20}), presented below.
\begin{equation}
\begin{aligned}
&
\frac{dn}{dy}(Pb+Pb) = A_{_{AA}}(K_{_{AA}},EnergySharing_{_{AA}}) \cdot B_M \cdot (\sqrt{s} -2m_\text{p}) \cdot F_{_{AA}}(y) \; ,
\end{aligned}
\label{eq15}
\end{equation}
\vspace*{-0.9cm}

\begin{equation}
\begin{aligned}
&
\frac{dn}{dy}(p+p) = A_{pp}(K_{pp},EnergySharing_{pp}) \cdot B_M \cdot (\sqrt{s} - 2m_\text{p}) \cdot F_{_{AA}}(y)  \; \; \; , \\
\end{aligned}
\label{eq16}
\end{equation}

\begin{equation}
\begin{aligned}
&
\frac{dn}{dy}(p+p) = B_M \cdot \tilde{\tilde{A}} \cdot EnergySharing_{pp} \cdot 2E_{inel} \cdot F_{_{AA}}(y)  \; \; \; , \\
\end{aligned}
\label{eq17}
\end{equation}

\begin{equation}
\begin{aligned}
&
\frac{dn}{dy}(Pb+Pb) = B_M \cdot \tilde{\tilde{A}} \cdot EnergySharing_{_{AA}} \cdot 2E_{inel} \cdot F_{_{AA}}(y)  \; \; \; , \\
\end{aligned}
\label{eq18}
\end{equation}

\begin{equation}
\begin{aligned}
&
A_{pp}(K_{pp},EnergySharing_{pp}) = \tilde{\tilde{A}}\cdot EnergySharing_{pp}\cdot K_{pp}  \; \; \; , \\
\end{aligned}
\label{eq19}
\end{equation}

\begin{equation}
\begin{aligned}
&
A_{_{AA}}(K_{AA},EnergySharing_{_{AA}}) = \tilde{\tilde{A}}\cdot EnergySharing_{_{AA}}\cdot K_{_{AA}}  \; \; \; . \\
\end{aligned}
\label{eq20}
\end{equation}
%
%
In the formulae above, 
the normalization of the pion $\frac{dn}{dy}$ distribution is 
now
a function of 
both the
baryon inelasticity $K$ and of 
the
sharing of the available inelastic energy.
The quantity
$EnergySharing$ describes the part of this available energy spent on pions.
$\tilde{\tilde{A}}$ is a constant factor.
Following section~\ref{IV.3}, 
$EnergySharing$  is respectively:
\begin{equation}
\begin{aligned}
&
EnergySharing_{pp} \approx 1/(1 + 0.13378)\; \; \; , ~~~~\text{from Eq.~(\ref{YPPK}), for p+p collisions},\\
&
EnergySharing_{_{AA}} \approx 1/(1 + 0.26333)\; \; \; , ~~~\text{ from Eq.~(\ref{YPBPBK}), for Pb+Pb collisions}.
\end{aligned}
\label{eq21}
\end{equation}
%
Thus the normalization ratio for
the two distributions (\ref{eq16}) and (\ref{eq15})
is
%
\begin{equation}
\begin{aligned}
&
\frac{A_{pp}}{A_{_{AA}}}=
 \frac{EnergySharing_{pp}\cdot K_{pp}}{EnergySharing_{_{AA}}\cdot K_{_{AA}}}= 0.781  \; \; \; ,
\end{aligned}
\label{eq22}
\end{equation}
which is a direct reflection of Eq.~(\ref{XUXU}).

Let us underline that the normalization ratio of 0.781 given above is the {\em only difference} between the 
function with which we approximated the $\frac{dn}{dy}$ distribution   of negative pions in
p+p reactions 
              (Eq.~(\ref{eq4}), consequently~(\ref{eq6}) 
and (\ref{eq16})) and 
the one
which we obtained
for 
the
fire-streak in
Pb+Pb collisions (Eq.~(\ref{fragmentation}), consequently~(\ref{eq5}) and~(\ref{eq15})). 
This value 
of 0.781
has been deduced solely from our estimates of the energy balance between pion, kaon and baryon emission in p+p and in Pb+Pb events.
These latter estimates have been
obtained directly from interpolated experimental data on $\pi^\pm$, $K^\pm$, net $p$, and $n$ production,
with only a minimal set of basic assumptions in sections~\ref{IV.1},~\ref{IV.2}, and~\ref{IV.3}.  


The value of 0.781 is now to be compared with the factor 
 0.812
which we found from the 
comparison 
of our function $f(y)$
to the 
isospin-corrected
$\pi^-$ rapidity distribution
in Fig.~\ref{fig5},
and subsequently stated in Eq.~(\ref{eq4.5}). 
This gives us a 4\% agreement which we consider as 
very good, 
account 
taken 
of 
the uncertainties inherent to our 
%
%
study.\footnote{We note that the latter include both our assumptions and approximations as well as the uncertainties of the experimental p+p and Pb+Pb data which we used. For instance, the systematic errors of the experimental pion $dn/dy$ yields in Pb+Pb collisions reach 5-10\% depending on centrality~\cite{2.5na49}.} 
%
%



From the above, we find it justified to conclude that the agreement of shapes shown in Fig.~\ref{fig5} can now be re-interpreted as a {\em full overall consistency} of the experimental $\pi^-$ rapidity distribution in p+p collisions with the {\em absolutely normalized} fire-streak fragmentation function. Indeed, directly from Eqs.~(\ref{eq4}) and~(\ref{eq22}), the following 
becomes true:
\hspace*{-0.0cm}
\begin{center}
{Experimental $\pi^-$ rapidity distribution in p+p collisions \\
$\approx$~ 
fire streak fragmentation function
into $\pi^-$}
\end{center}
\vspace*{-1cm}
\begin{equation}
\label{eqtext}
\end{equation}
\vspace*{-0.1cm}

\hspace*{-0.6cm}
- up to the 4\% accuracy in normalization mentioned above.\
This occurs
once 
the correction for 
isospin effects 
is
taken into account (Eq.~(\ref{eq2})), and another correction for strangeness enhancement and baryon inelasticity differences between p+p and Pb+Pb reactions 
is 
included in
the 
comparison
(Eq.~(\ref{eq22})). 
We will further discuss these issues in section~\ref{V}.


\subsection{Comment on Eq.~(\ref{eq15}) }
\label{IV.5}

For completeness and clarity of the discussion made in section~\ref{V}, below we rewrite formula~(\ref{eq15}) in the form evident from Eq.~(\ref{eq20}):
\begin{equation}
\frac{dn}{dy}= 
\tilde{\tilde{A}} 
\; \cdot \; EnergySharing 
\; \cdot \; K 
\; \cdot \; B_M 
\; \cdot \; (\sqrt{s} -2m_\text{p}) 
\; \cdot \; 
\exp\left( - \frac{[       y^2 + \epsilon^2]^\frac{r}{2}}{r \cdot \sigma_{y}^{r}} ~ \right)\; \; .
\label{eq23}
\end{equation}
In the above we dropped all the reaction-specific 
indices and wrote explicitly the shape factor introduced in Eq.~(\ref{eq5}).
The parameters $\epsilon$, $\sigma_y$, and $r$ are obtained from the fit to Pb+Pb collisions (section~\ref{II}), and 
  $\tilde{\tilde{A}}=0.0907$ 
from Eq.~(\ref{eq20}). 
The formula~(\ref{eq23}) gives our fire-streak fragmentation function 
in central Pb+Pb collisions, at $b=0$. 
After the correction for 
strangeness suppression 
in p+p relative to Pb+Pb collisions
and for the difference in baryon inelasticity 
(parametrized respectively by $EnergySharing$ and $K$), the same formula gives
the blue curve which approximately describes the isospin corrected p+p data points in Fig.~\ref{fig5} (within 4\% accuracy as discussed in section~\ref{IV.4}).

\vspace*{-0.2cm}
\section{Discussion}
\label{V}
\vspace*{-0.2cm}

In this section we will attempt to draw the conclusions from the findings made in the present study, partially in the context of these made in our earlier work~\cite{1}.

Our initial 
concept~\cite{1},
with some similarity to the fire-streak picture~\cite{2}, was 
introduced in order to explain the role of geometry and 
local energy-momentum conservation in the centrality dependence of Pb+Pb collisions at SPS energies. Simultaneously, our work~\cite{1} was inspired by, and meant to explain, our observations from spectator-induced electromagnetic effects on $\pi^+/\pi^-$ ratios and directed flow in heavy ion collisions~\cite{twospec07,Rybicki_v1,Rybicki2015,wpcf}, indicating that pions at higher rapidity 
are
produced closer to the spectator system as it is suggested by Fig.~\ref{fig1}.

The result was that the full centrality dependence of pion rapidity distributions and total pion yields could be understood from three elements: (a) collision geometry (b) local energy-momentum conservation, and (c) our simple fire-streak fragmentation function, producing pions proportionally to the available energy (Eq.~(\ref{fragmentation})).

With the present work, however, a new element appears in the picture which is the (exact or approximate) consistency of the isospin corrected experimental $\pi^-$ rapidity distribution in p+p reactions with the 
fire-streak fragmentation function, as shown in Fig.~\ref{fig5} and stated in section~\ref{IV.4}.
This
consistency emerges {\em only} when the normalization of the latter 
is corrected for the change in baryon inelasticity and the strangeness enhancement between p+p and Pb+Pb collisions. This brings specific implications,
some of which we will point below.
%
\vspace*{-0.1cm}

\subsection{Pion rapidity spectra}
\vspace*{-0.1cm}

In the present study, one component of our successful description of pion production in
Pb+Pb reactions from Ref.~\cite{1} - the fire streak fragmentation function - appears ``available'' 
in 
p+p collisions
once the effects of baryon inelasticity and strangeness suppression are taken into account.
 Thus one
can think of the following simple 
``prescription''
to follow in order to describe, or parametrize, 
the centrality dependence of pion rapidity distributions and their total yields in Pb+Pb reactions, starting from p+p collisions:
\vspace*{0.2cm}
\begin{center}

pion $dn/dy$ distribution in p+p collisions (Fig.~\ref{fig3}) 

{\large $\Downarrow$}

correction for isospin (Eq.{\small~(\ref{eq2})})

{\large $\Downarrow$}

isospin corrected pion $dn/dy$ distribution in p+p
(Eq.{\small~(\ref{eq23}))}

{\large $\Downarrow$}

correction for change in baryon inelasticity and strangeness enhancement 
(Eq.{\small~(\ref{eq23}))}

{\large $\Downarrow$}

fire-streak fragmentation function in Pb+Pb (Eq.{\small~(\ref{eq23}))}

{\large $\Downarrow$}

collision geometry {\Large +} local $E-\vec{p}$ conservation (\cite{1}, Fig.~\ref{fig1})

{\large $\Downarrow$}

pion $dn/dy$ distributions in Pb+Pb as a function of centrality (Fig.~\ref{fig2})\\

\end{center}
\vspace*{0.3cm}

We underline that the scheme above may be followed both ``down'' and ``up''. For instance, our study made in Ref.~\cite{1} supplemented by the present analysis, follows it ``up'' from the centrality dependence of 
the
Pb+Pb 
reactions
up to the pion spectrum in p+p collisions. 
%
%
%
%
The prescription established above will keep track of the whole shape evolution of the $dn/dy$ distribution from p+p through peripheral up to central Pb+Pb collisions, and of the relative increase of pion multiplicity as a function of 
decreasing
impact parameter of the Pb+Pb collision. 
In our view,
this
``correspondence'' between 
rapidity distributions in
p+p and Pb+Pb interactions established by our 
prescription
brings additional support
to
our
simple
picture of the longitudinal evolution of the 
Pb+Pb 
system.
In this picture,
finite size volumes of deconfined
primordial
matter initially move following local energy-momentum conservation, and a number of mechanisms resulting in production of final state particles in Pb+Pb collisions
(dressing up of quarks into hadrons, etc) preserve 
some degree of 
similarity to p+p reactions. 

\vspace*{-0.1cm}

\subsection{Differences between p+p and Pb+Pb collisions}
\vspace*{-0.1cm}

As a continuation of our
paper~\cite{1}, the present work is aimed at pointing out possible common points and similarities in pion rapidity distributions for the two reactions. Its limitations should also be pointed out. Evidently, 
our work 
does not genuinely 
``explain'' 
strangeness enhancement nor the changes in inelasticity $K$ between proton-proton and nucleus-nucleus collisions. Both of these we had to estimate from experimental data in section~\ref{IV} for the purpose of formula~(\ref{eq23}). 
Specifically, our ``correction'' for strangeness suppression in p+p or strangeness enhancement in Pb+Pb reactions, introduced by the estimated quantity $EnergySharing$ in Eq.~(\ref{eq23}), is in fact a simple ``translation'' of enhanced strange particle yields into the overall energy balance of particle production.
The origin of this correction - the enhanced abundance of strange quarks in the deconfined matter produced in Pb+Pb collisions - is an independent dynamical phenomenon explained elsewhere~\cite{rafelski}. It evidently modifies the overall energy balance in particle emission but it is only parametrized in our study.
As such, no claim can be made about bulk properties of heavy ion collisions being predictable 
{solely}
from p+p reactions on the basis of the present work.

{Also, in our view, our results do not point towards the applicability of the geometrical 
picture of many fire-streaks, as drawn in Fig.~\ref{fig2}, to proton-proton reactions.} This is in 
contrast to our work on pion $dn/dy$ distributions in Pb+Pb collisions~\cite{1}. The fact that consistency can be found between the experimental pion rapidity distribution in p+p collisions and the fragmentation function of the {\em single} fire-streak, rather than a {\em sum} of fire-streaks, 
{would suggest
a difference between the two reactions.
While Pb+Pb data can be described by a superposition of many independent fire-streaks, only a single fire-streak would be formed in the p+p collision.}



\vspace*{-0.2cm}
\section{Summary}
\label{VI}
\vspace*{-0.2cm}

In the present paper we 
investigated to which extent the phenomenological rapidity distribution
of pions from the fire-streak in Pb+Pb collisions, extracted recently,
is similar to the pion rapidity distribution in p+p collisions.
With no tuning nor adjustment to experimental data, our single fire-streak pion $\frac{dn}{dy}$ distribution
obtained from Pb+Pb reactions reproduced the shape of the experimental pion rapidity spectrum in p+p interactions at the same energy. 
Isospin differences between Pb+Pb and p+p collisions have been taken into account.
The 
absolute normalization of pion spectra between the two reactions could be fully (up to 4\% precision) explained by changes in the energy balance induced by
baryon stopping and strangeness enhancement phenomena.

From the above we conclude that once the above 
phenomena
are taken into account, and the influence of Pb+Pb reaction geometry as well as local energy-momentum conservation are properly considered, 
an
interesting
 correspondence 
emerges
between absolutely normalized pion rapidity spectra in inelastic p+p 
collisions 
and pion rapidity distributions in centrality selected Pb+Pb reactions.\\


{\bf Acknowledgments}\\

We gratefully thank Adam Bzdak for pointing to us the importance of the extension of our study to p+p reactions.
We acknowledge the work of Hans~Gerhard~Fischer on the release and especially the precise numerical interpolation of NA49 proton+proton data which allowed a model-independent calculation of the energy balance in p+p collisions. 
We are indebted to Jan Rafelski for his remarks on the fire-streak model,
{and to our referees for very valuable comments and constructive criticism.}
This work was supported by the National Science Centre, Poland (grant number
2014/14/E/ST2/00018).



\newpage
\vspace*{2cm}
{\bf APPENDIX A. THE ENERGY BALANCE IN P+P REACTIONS AT SPS ENERGIES}\\

In the following we cross-check the overall energy balance in p+p reactions at the top SPS energy ($\sqrt{s}$=17.27~GeV) as emerging from our considerations made in section~\ref{IV}. Following the approximation made therein in Eq.~(\ref{YPP}) we assume that the energy $E_{inel}$ lost by the incoming baryon is spent uniquely on final state pion and kaon production. 
This means that we neglect the production of baryon-antibaryon pairs as well as other less abundant particles.  
Under this assumption the partition of energy in the final state writes:
\begin{equation}
{\sqrt{s}}
\approx \mathrm{(net~baryon~energy)}+\mathrm{(pion~energy)}+\mathrm{(kaon~energy)} \; \; \; \; \; ,\\
\label{BAL1}
\end{equation}
where each of the three terms 
 corresponds
to the average summed energy of all the net baryons, pions and kaons in the inelastic p+p event. Relating this to the baryon inelasticity $K$ introduced in Eq.~(\ref{k}) we obtain:
\begin{equation}
{\sqrt{s}}
\approx 2m_\text{p} + (\sqrt{s}-2m_\text{p})\cdot (1-K) 
+\mathrm{(pion~energy)}+\mathrm{(kaon~energy)} \; \; \; \; \; ,
\label{BAL2}
\end{equation}
where $m_\text{p}$ is the proton mass and the difference between the latter and the neutron mass is neglected. We assume 
 $K=0.547$ 
which we obtained for summed net protons and net neutrons in section~\ref{IV.1}; thus, we neglect the small changes in the 
net 
baryon
term of Eq.~(\ref{BAL1}), possibly induced by the presence of $\Lambda$ as well as other baryons in the final state. From Eq.~(\ref{XPP}) we have:
\begin{equation}
\begin{aligned}
&
\mathrm{(pion~energy)}=E(pp\rightarrow\pi)=6862~\mathrm{MeV} \; \; \; ,\\
&
\mathrm{(kaon~energy)}=
E(pp\rightarrow K^{+})+
E(pp\rightarrow K^{-})+
E(pp\rightarrow K^{0\overline{0}})=918~\mathrm{MeV} \; \; \; ,
\end{aligned}
\label{BAL3}
\end{equation}
and Eq.~(\ref{BAL2}) writes:
\begin{equation}
{\sqrt{s}}
\approx
2\cdot 0.938 +
 15.394 \cdot (1-0.547) +
6.862 +
0.918
=
 16.629~\mathrm{GeV} \; \; \; .
\label{BAL4}
\end{equation}
In comparison to the original value of $\sqrt{s}=17.27$~GeV, this gives us the 
 3.7\%
agreement 
mentioned
in section~\ref{IV.3}, which we consider good enough 
taken the accuracy of
the present work.

It is interesting to consider the impact of other particles, neglected in the present study, on the overall energy balance in p+p reactions. While a complete study is beyond the scope of this paper, we note that to evaluate this impact is most straight-forward for the contribution of non-strange baryon-antibaryon pairs, that is, pair produced $p$, $\bar{p}$, $n$, and $\bar{n}$. For antiprotons, precise wide-acceptance double-differential $(x_F,p_T)$ distributions are available in p+p collisions from the NA49 experiment~\cite{pprot}, including a precise numerical interpolation~\cite{spshadrons} as it was the case for the other particles discussed in section~\ref{IV}. Thus we can apply formula~(\ref{e_av_i}) assuming 
$m_i=m_\mathrm{\bar{p}}$ to estimate the mean energy of an antiproton produced in inclusive inelastic p+p collisions at $\sqrt{s}=17.27$~GeV. We obtain:
\begin{equation}
\langle E_{i}\rangle
 ~=~ 
\langle E_{\mathrm{\bar{p}}}\rangle
 ~=~ 
1451~\mathrm{MeV}
\; \; \; \; \; .
\label{e_av_pbar}
\end{equation} 
Subsequently, account taken of the published average multiplicity 
of
0.0386 antiprotons per inclusive inelastic p+p event~\cite{pprot}, we get the average energy spent for antiproton production:  
%
%
\begin{equation}
E(pp\rightarrow\overline{p})  =  0.0386 \cdot 1451 = 56~\mathrm{MeV} \; \; \; .\\
\label{XPPpbar}
\end{equation}
Following the considerations made in~\cite{pprot}, we multiply the above by 1.66 in order to obtain the average energy spent on antineutron production. Finally we multiply the summed antiproton+antineutron contribution by two in order to get the total average energy which an inelastic p+p collision spends on pair-produced protons, neutrons, antiprotons and antineutrons:
\begin{equation}
E(pp\rightarrow\text{non-strange,~pair-produced}~B\text{~and~}\overline{B})  =  2 \cdot (1 + 1.66) \cdot 56 = 298~\mathrm{MeV} \; \; \; .\\
\label{XPPpbar}
\end{equation}
Adding the above value to the right side of Eq.~(\ref{BAL1}) we obtain $\sqrt{s}\approx 16.927$~GeV in Eq.~(\ref{BAL4}), which gives an agreement within 2\% with the original value of 17.27~GeV. Thus already the inclusion of non-strange baryon and antibaryon pair production improves the accuracy of our energy balance by a factor of two. We note that a 2\% accuracy seems excellent to us, and emphasizes the 
quality of the published experimental data on p+p collisions at SPS energies which we used in this study~\cite{ppaper,pprot,kaonpaper}.

\newpage
{\bf APPENDIX B. ENERGY DEPENDENCE OF THE FRAGMENTATION FUNCTION}\\

\begin{figure}[b]
\begin{center}
\vspace*{-0.8cm}
\includegraphics[width=12cm,height=15cm]{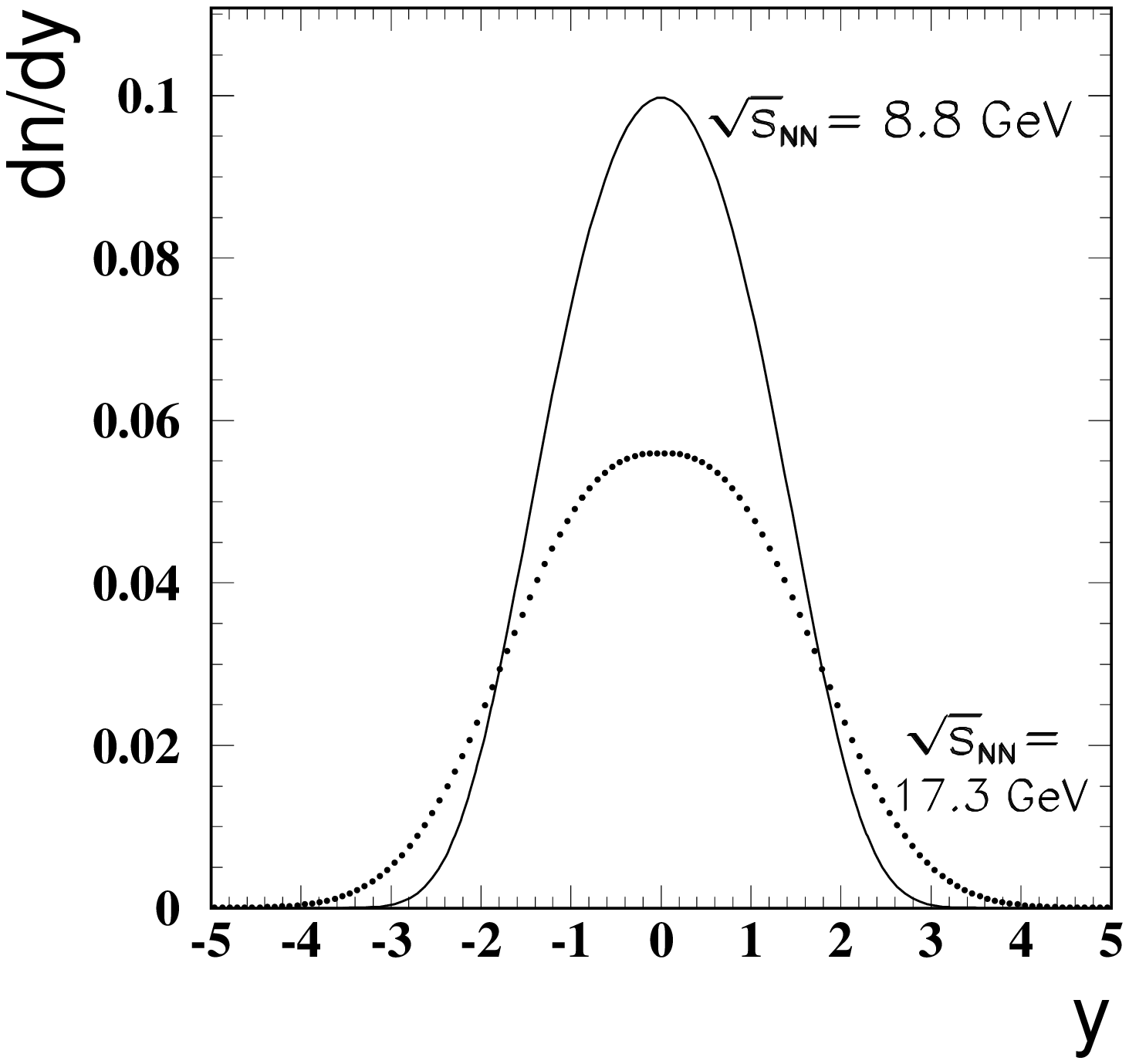}
\vspace*{-7.5cm}
{\caption\small Comparison of single fire-streak fragmentation functions used for the description of $\pi^-$ rapidity distributions in Pb+Pb collisions at $\sqrt{s_{NN}}=8.8$~GeV (solid) and at $\sqrt{s_{NN}}=17.3$~GeV (dotted). The two presented functions are given by Eq.~(\ref{fragmentation}) with  
$(E^*_\text{s}-m_\text{s})\equiv 1$~GeV.
The numerical values of the function parameters are given in the text.\label{Xfr}}   
\end{center}
\end{figure}

While this paper was in principle not devoted to the energy dependence of nucleus-nucleus collisions,
the completeness of the discussion requires that we comment on the comparison between Fig.~\ref{fig2}(b) and Fig.~\ref{fig2-40} made in section~\ref{intro}. Following section~\ref{II}, the numerical parameters of the single fire-streak fragmentation function providing the best description of negative pion spectra in Pb+Pb collisions at $\sqrt{s_{NN}}=17.3$~GeV (see Fig.~\ref{fig2} and Eq.~(\ref{fragmentation})) are $A=0.05598$, $\sigma_y=1.475$, $r=2.55$, and $\epsilon=0.01$. For the lower collision energy of $\sqrt{s_{NN}}=8.8$~GeV (see Fig.~\ref{fig2-40}), the corresponding single fire-streak fragmentation function obeys the parametrization given by Eq.~(\ref{fragmentation}), but with different numerical parameters: $A=0.173$, $\sigma_y=1.800$, $r=4.60$, and $\epsilon=2.203$. 
At the present moment we do not attribute much 
physical sense to the changes 
of each given parameter taken separately, as we anyway consider the functional shape given by the exponent in Eq.~(\ref{fragmentation})
as a purely effective approximation of a complex non-perturbative process (see also the discussion made in section~\ref{III}, item(3)). 
What we consider
important is that at both 
collision energies the fragmentation function keeps the proportionality of the number of produced pions to the available energy,
$(E^*_\text{s}-m_\text{s})$
 in Eq.~(\ref{fragmentation}). This supports energy-momentum conservation as the main basis for our model, and its connection to p+p collisions which we formulated in sections~\ref{II}-\ref{V}.

On the other hand, it is worthwhile to perform a direct comparison of the two fragmentation functions. This is presented in Fig.~\ref{Xfr}. Both functions are
taken assuming the same available energy in the fire-streak, that is, 
$(E^*_\text{s}-m_\text{s})\equiv 1$~GeV
 in Eq.~(\ref{fragmentation}).
Thus our comparison reflects both the change in the shape of the pion rapidity distribution and the change in the number of pions produced {\em per one GeV of available energy}, as a function of $\sqrt{s_{NN}}$.

In our 
view,
a 
consistent picture emerges from Fig.~\ref{Xfr}. With increasing collision energy, the fragmentation function broadens in rapidity, while its total integral decreases visibly. 
This nicely reflects 
the phenomenon of 
broadening of rapidity spectra of produced particles with increasing reaction energy, as well as the slower than linear increase of their total 
multiplicity as a function of $\sqrt{s}$.

\vspace*{2cm}


\end{document}